\documentstyle[aps]{revtex}

\begin{document}
\title{Noise induced transitions in semiclassical cosmology}
\author{{\footnotesize Esteban Calzetta$^{\dag }$ and Enric Verdaguer$^{\S }$}}
\address{$\dagger $Instituto de Astronom\'{\i}a y F\'{\i}sica del\\
Espacio (IAFE) and Departamento de F\'{\i}sica,\\
Universidad de\\
Buenos Aires,\\
Ciudad Universitaria,\\
1428 Buenos Aires, Argentina\\
\S\ Departament de F\'{\i}sica Fonamental and Institut de F\'{\i}sica\\
d'Altes Energies (IFAE),\\
Universitat de Barcelona, Av. Diagonal 647, E-08028 Barcelona, Spain}
\maketitle

\begin{abstract}
A semiclassical cosmological model is considered which consists of a closed
Friedmann-Robertson-Walker in the presence of a cosmological constant, which
mimics the effect of an inflaton field, and a massless, non-conformally
coupled quantum scalar field. We show that the back-reaction of the quantum
field, which consists basically of a non local term due to gravitational
particle creation and a noise term induced by the quantum fluctuations of
the field, are able to drive the cosmological scale factor over the barrier
of the classical potential so that if the universe starts near zero scale
factor (initial singularity) it can make the transition to an exponentially
expanding de Sitter phase. We compute the probability of this transition and
it turns out to be comparable with the probability that the universe tunnels
from ``nothing" into an inflationary stage in quantum cosmology. This
suggests that in the presence of matter fields the back-reaction on the
spacetime should not be neglected in quantum cosmology.
\end{abstract}

\section{Introduction}

\label{sec:intro}

A possible scenario for the creation of an inflationary universe is provided
by cosmological models in which the universe is created by quantum tunneling
from ``nothing'' into a de Sitter space. This creation is either based on an
instanton solution or in a wave function solution which describes the
tunneling in a simple minisuperspace model of quantum cosmology \cite
{Vilenkin 1982-1984,Hartle and Hawking}.

In the inflationary context one of the simplest cosmological models one may
construct is a closed Friedmann-Robertson-Walker (FRW) model with a
cosmological constant. The cosmological constant is introduced to reproduce
the effect of the inflation field at a stationary point of the inflaton
potential \cite{Vilenkin 1982-1984}. The dynamics of this universe is
described by a potential with a barrier which separates the region where the
scale factor of the universe is zero, where the potential has a local
minimum, from the region where the universe scale factor grows
exponentially, the de Sitter or inflationary phase. The classical dynamics
of this homogeneous and isotropic model is thus very simple: the universe
either stays in the minimum of the potential or it inflates.

The classical dynamics of the preinflationary era in such cosmological
models may be quite complicated, however, if one introduces anisotropies,
inhomogeneities or other fields. Thus, for instance, all anisotropic Bianchi
models, except Bianchi IX, are bound to inflate in the presence of a
cosmological constant \cite{Wald 1983}. Also in the previous model but with
an inhomogeneous scalar radiation field the universe may get around the
barrier \cite{Calzetta and El Hasi 1995} and emerge into the inflationary
stage even if initially it was not.

The emergence of an inflationary stage of the universe also seems to be
aided by semiclassical effects such as particle creation which enhances the
radiation energy density of the preinflationary era and thus enlarges the
set of inflating initial conditions \cite{Calzetta 1991,Calzetta and
Sakellariadou 1992}.

In this paper we consider a semiclassical model consisting of a closed FRW
cosmology with a cosmological constant in the presence of a quantum massless
scalar field. This quantum field may be seen as linear perturbations of the
inflaton field at its stationary point or as some other independent linear
field. Because the field is free the semiclassical theory is one loop exact. The
expectation value in a quantum state of the stress-energy tensor of this scalar
field influences by back-reaction the dynamics of the cosmological scale factor.
There are here two main effects at play: on the one hand, since the field is not
conformaly coupled particle creation will occur and, on the other hand, the
quantum fluctuations of this stress-energy tensor induce stochastic classical
fluctuations in the scale factor \cite {CV97,CCV97}. Thus the cosmological scale
factor is subject to a history dependent term due to gravitational particle
creation and also to noise due to these quantum fluctuations. We examine the
possibility that a universe starting near the local minimum may cross the
barrier and emerge into the inflationary region by the back-reaction of the
quantum field on the scale factor. This is, in some sense, the semiclassical
version of tunneling from nothing in quantum cosmology.

It is important to stress the difference between this calculation and the
usual approach to quantum tunneling. The usual approach \cite
{Coleman,Affleck,Vilenkin 1982-1984,Garriga 1994,Hawking and Turok} begins
with the calculation of an instanton or tunneling solution, which is a
solution to the Euclidean classical (or sometimes semiclassical, see \cite
{Barvinsky}) equations of motion. Because of symmetry, the scalar field is
set to zero from the start. Its effect, if at all, is considered as a
contribution to the prefactor of the tunneling amplitude 
\cite{Garriga 1994},
which is usually computed to one loop accuracy in the test field
approximation. The effect of dissipation \cite{Caldeira 1983} or even of
particle creation \cite{Rubakov 1984} on quantum tunneling has been
considered in some quantum mechanical systems but it ought to be noticed
that to this date the effects of stress-energy fluctuations on the tunneling
amplitude has not been considered in the literature, to the best of our
knowledge. Even when the instanton is sought as a solution to semiclassical
equations \cite{Barvinsky} this is done under approximations that
effectively downplay the role of particle creation, and back-reaction
fluctuations are not considered at all.

To underlie that the mechanism for barrier penetration to be investigated
here is a different physical process than that computed from instantons in
the test field approximation, we have chosen to ignore the quantum aspects
of the gravitational field, so that in the absense of back reaction
fluctuations the tunneling rate would be zero. From the point of view of the
usual approximation, it could be said that our calculation amounts to a
nonperturbative calculation of the tunneling amplitude, since the key
element is that we go beyond the test field approximation, and consider the
full effect of back reaction on the universe.

At least in principle, it ought to be possible to combine both the usual and
our approach. The whole scheme would ressemble the derivation of the
Hu-Paz-Zhang equations \cite{HPZ,Halliwell}, once the subtleties of quantum
cosmological path integrals are factored in \cite{Halliwell 1988}.

In this paper, we follow the methodology of Langer's classic paper 
\cite{Langer}, namely, we shall consider an ensemble
of Universes whose evolution is rendered stationary by the device that, every
time a member of the ensemble escapes the barrier it is captured and
reemitted within the barrier. This fictitious stationary solution has a nonzero
flux accross the barrier, and the activation
probability is derived from this flux.

Since semiclassical cosmology distinguishes a particular time (that when the 
quantum to classical transition takes place), it
is meaningful to ask whether the stationary solution is relevant to the 
behavior of a solution with arbitrary initial data at the 
``absolute zero of time''. The answer is that the stationary solution is 
indeed relevant, because the relaxation time which
brings an arbitrary solution to the steady one is exponentially shorter than 
the time it takes to escape the barrier. We discuss 
this issue in detail in Appendix F.

The fact itself of  assuming a semiclassical theory, i.e., where no 
gravitational fluctuations are included, indicates that our model
must be invalid very close to the cosmological singularity. Therefore, we are 
forced to assume that some mechanism forces the universe
to avoid this region, while being too weak to affect significatively the 
behavior of larger universes. For example, if we take the
cosmological constant, in natural units, to be about $10^{-12}$ (which 
corresponds to GUT scale inflation), then the presence of 
classical radiation with an energy density of order one (while the amount 
necessary to avoid recollapse in the classical theory is 
$10^{12}$), would be sufficient. A  more sophisticated possibility would be to 
appeal to some quantum gravitational effect, which could be
as simple as Heisenberg's uncertainty principle, to make it impossible for the
universe to linger for long times too close to the singularity.

Even with this simple setting, it is impossible to make progress without 
further simplifications, and we would like to give here
a summary of the most significative ones. The most basic simplifying 
assumption is that the deviation from conformal coupling,
measured by the parameter $\nu$ to be introduced below (see Eq. (\ref{2})), is
small.  This will allow us to set up the problem as a
perturbative expansion in $\nu$, whereby we shall stick to the lowest 
nontrivial order, namely $O(\nu^2)$. Of course, the quantity of
highest interest, the escape probability itself, will turn out to be 
nonperturbative in $\nu$; however, our procedure ought to capture
its leading behavior.

Even to second order in $\nu$, the Closed Time Path (CTP) effective action,
whose  variation yields the semiclassical equations for 
the universe scale factor, involves the calculation of several kernels. We have
formal  exact expressions for these kernels, but the results
are too involved for further manipulation. This suggests a second 
simplification, namely, to substitute the exact kernels for 
their analogs as computed in an spatially flat universe with the same scale
factor. Technically, this amounts to making a continuous 
approximation in the mode decomposition of the field. This is clearly 
justified when the separation between the frequencies for
different modes is small, for example, as compared with the characteristic 
rate of the universe espansion. This condition holds
for most orbits within the barrier, excepting maybe those where the universe 
never grows much larger than Planck's scales, a case which we shall not 
discuss, for the reasons given above.

The semiclassical evolution equations emerging from the CTP 
effective action differ from the usual Einstein equations in
three main respects: 1) the polarization of the scalar field vacuum induces 
an effective potential, beyond the  usual terms  associated to spatial curvature
and the cosmological  constant; also the gravitational constants are
renormalized by quantum fluctuations; 2) there appears a memory dependent term,
asociated to the stress-energy of particles created along the 
evolution; and 3) there appears a stochastic term associated to the quantum 
fluctuations of the scalar field. We shall focus our
attention in the last two aspects, neglecting the one loop effective 
gravitational potential. It ought to be noted that, lacking a 
theory of what the bare potential is exactly like, the semiclassical theory does
not uniquely  determine the renormalized potential either. Moreover, the
presence of stochasticity and memory are aspects where the semiclassical 
physics is qualitatively different from the classical one, not
so for the modified effective potential. In any case, these corrections are 
very small unless very close to the cosmological singularity (where in any case
the one loop approximation is unreliable,  as implied by the 
logarithmic divergence of the quantum corrections). So,
assuming again that some mechanism  will make
it impossible for the universe to stay very close to the 
singularity, the neglect of the renormalized potential is justified.

Even after the neglect of the renormalized potential, the equations deriving
from the CTP effective action are higher than second order,
therefore do not admit a Cauchy problem in the usual terms, and also lead
to possibly unphysical solutions. In order to reduce them to second order
equations, and to ensure that the solutions obtained are physical, it is
necessary to implement an order reduction procedure as discussed by
several authors \cite{Flanagan and Wald 1996}. 
This order reduction means that higher
derivatives are expressed in terms of lower ones as required by the classical
equations of motion. In this spirit, in the memory term, we substitute the
history of a given state of the universe by the classical trajectory leading
up to the same endpoint. Because the classical trajectory is determined by
this endpoint, in practice this reduces the equations of motion to a local form,
although no longer Hamiltonian.

The equations of motion for the model, after all these simplifications have
been carried out, have the property that they do not become singular when
the universe scale factor vanishes. As a consequence,  the universe goes
accross the cosmic singularity and emerges in a new ``cosmic cycle''. Because
the escape time is generally much larger than the recollapse time, we may
expect that this will happen many times in the evolution of a single
trajectory. For this reason, our model describes a cyclic universe, being 
created and destroyed many times (but keeping the memory of the
total amount of radiation and extrinsic curvature at the end of the previous 
cycle), and eventually escaping from this fate to become an inflationary
universe. It should be noted that this does not detract from the rigor of
our derivations, since it is after all a feature of the mathematical model,
it being a matter of opinion whether it affects the application of our studies
to the physical universe. For comparison we have studied a different, also
mathematically consistent, model in which the universe undergoes a single cosmic
cycle and obtain similar results (see Appendix G).

After this enumeration of the main simplifying assumptions to be made below,
let us briefly review what we actually do. Our first concern is to derive the
semiclassical equations of motion for the
cosmological scale factor, by means of the CTP effective action. The imaginary
terms in this action can be shown to carry the information about the stochastic
noises which simulate the effect on the geometry of 
quantum fluctuations of the matter field \cite{CH94,hargell,Gleiser and Ramos
1994,CH95,HuSin,CV96,eft}.
After this noises have been identified, the semiclassical equation is upgraded
to a Langevin equation.

We then transform this Langevin equation 
into a Fokker-Planck equation, and further simplify it by averaging along
classical trajectories. In this way, we find an evolution equation for the
probability density of the universe being placed  within a given classical
trajectory. The actual universe jumps between classical trajectories, as it
is subject to the non Hamiltonian nonlocal terms and forcing from the random
noises. Finding the above equation of evolution requires a
careful analysis of both effects.

Finally, we investigate the steady solutions of this equation, and derive the
escape probability therein. Again we are forced to consider the problem of
very small universes, as the nontrivial steady solutions are nonintegrable
in this limit. However,  the solutions to the Wheeler-DeWitt
equation associated to  our model, which in this limit is essentially the
Schr\"odinger operator for an harmonic oscillator, shows no singular behavior
for small universes. Thus we shall assume that this divergence will be
cured in a more complete model, and
accept the nontrivial solution as physical.

The main conclusion of this paper is that the probability that the universe
will be carried over the barrier by the sheer effect of random forcing from
matter stress-energy fluctuations is comparable to the tunneling
probability computed from gravitational instantons. This effect demonstrates
the relevance of quantum fluctuations in the early evolution of the
universe.

Besides its relevance to the birth of the universe as a whole, this result
also may be used to estimate the probability of the creation of inflationary
bubbles within a larger universe. We shall report on this issue in a further
communication.

The plan of the paper is the following. In section \ref{section2} we compute
the effective action for the cosmological scale factor and derive the
stochastic semiclassical back-reaction equation for such scale factor. In
section \ref{section3} we construct the Fokker-Planck equation for the
probability distribution function of the cosmological scale factor which
corresponds to the stochastic equation. In section \ref{section4} we use the
analogy with Kramers' problem to compute the
probability that the scale factor crosses the barrier and reaches the de
Sitter phase. In the concluding section \ref{section5} we compare our
results with the quantum tunneling probability. Some computational details are
included in the different sections of the Appendix. 

A short summary of this long
Appendix is the following: Appendix A gives some details of the
renormalization of the CTP effective action; Appendix B explains how to
handle the diffusion terms when the Fokker-Planck equation is constructed; in
Appendix C we formulate and discuss Kramers problem in action-angle variables;
the short Appendix D gives the exact classical  solutions for the cosmological
scale factor; in Appendix E the averaged diffusion and dissipation
coefficients for the averaged Fokker-Planck equation are derived; in
Appendix F the relaxation time is computed in detail; and finally in Appendix G
the calculation of the escape probability for the scale factor is made for a
model which undergoes a single cosmic cycle.

\section{Semiclassical effective action}
\label{section2}

In this section we compute the effective action for the scale factor of a
spatially closed FRW cosmological model, with a cosmological constant in the
presence of a quantum massless field coupled non-conformally to the spacetime
curvature. The semiclassical cosmological model we consider is described by
the spacetime metric, the classical source, which in this case is a
cosmological constant, and the quantum matter sources.

\subsection{Scalar fields in a closed universe}

The metric for a closed FRW model is given by,

\begin{equation}
ds^2=a^2(t)\left(-dt^2+ \tilde g_{ij}(x^k) dx^idx^j\right), \ \ \ \ i,j, k=
1,..., n-1,  \label{1}
\end{equation}
where $a(t)$ is the cosmological scale factor, $t$ is the conformal time,
and $\tilde g_{ij}(x^k)$ is the metric of an $(n-1)$-sphere of unit radius.
Since we will use dimensional regularization we work, for the time being, in 
$n$-dimensions.

Let us assume that we have a quantum scalar field $\Phi (x^\mu )$, where the
Greek indices run from $0$ to $n-1$. The classical action for this scalar
field in the spacetime background described by the above metric is
\begin{equation}
S_m=-\int dx^n\sqrt{-g}\left[ g^{\mu \nu }\partial _\mu \Phi ^{*}\partial
_\nu \Phi +\left( {\frac{n-2}{4(n-1)}}+\nu \right) R\Phi ^{*}\Phi \right] ,
\label{2}
\end{equation}
where $g_{00}=a^2$, $g_{0i}=0$, $g_{ij}=a^2\tilde g_{ij}$, $g$ is the metric
determinant, $\nu$ is a dimensionless parameter coupling the field to the
spacetime curvature ($\nu=0$ corresponds to conformal coupling), $R$ is the
curvature scalar which is given by
\begin{equation}
R=2(n-1){\frac{\ddot a}{a^3}} + (n-1)(n-4){\frac{\dot a^2}{a^4}} + 
(n-1)(n-2){\frac{1}{a^2}},  
\label{3}
\end{equation}
where an over dot means derivative with respect to conformal time $t$. Let
us now introduce a conformally related field $\Psi$
\begin{equation}
\Psi=\Phi a^{{\frac{n-2}{2}}},  \label{4}
\end{equation}
and the action $S_m$ becomes,
\begin{equation}
S_m=\int dtdx^1\dots dx^{n-1}\sqrt{\tilde g}\left[ \dot \Psi ^{*}\dot \Psi
-{\frac{(n-2)^2}4}\Psi ^{*}\Psi -\nu a^2R\Psi ^{*}\Psi +\Psi ^{*}\Delta
^{(n-1)}\Psi \right] ,  \label{5}
\end{equation}
where $\Delta^{(n-1)}$ is the $(n-1)$-Laplacian on the $(n-1)$-sphere,
\begin{equation}
\Delta^{(n-1)}\Psi\equiv {\frac{1}{\sqrt{\tilde g}}}\partial_i
\left( \sqrt{\tilde g}g^{ij}\partial_j\Psi\right).  
\label{7}
\end{equation}

Let us introduce the time dependent function $U(t)$
\begin{equation}
U(t)=-\nu a^2(t)R(t),  \label{8}
\end{equation}
and the d'Alambertian $\Box =-\partial_t^2+\Delta^{(n-1)}$ of the static
metric $\tilde ds^2=a^{-2}ds^2$. The action (\ref{5}) may be written then as,
\begin{equation}
S_m=\int dtdx^1\dots dx^{n-1}\sqrt{\tilde g}\left[ \Psi ^{*}\Box \Psi 
-{\frac{(n-2)^2}4}\Psi ^{*}\Psi +U(t)\Psi ^{*}\Psi \right] .  
\label{6}
\end{equation}

When $\nu=0$ this is the action of a scalar field $\Psi$ in a background of
constant curvature. The quantization of this field in that background is
trivial in the sense that a unique natural vacuum may be introduced, the
``in'' and ``out'' vacuum coincide and there is no particle creation \cite
{BD82}. This vacuum is, of course, conformally related to the physical
vacuum, see (\ref{4}). The time dependent function $U(t)$ will be considered
as an interaction term and will be treated perturbativelly. Thus we will
make perturbation theory with the parameter $\nu$ which we will assume small.

To carry on the quantization we will proceede by mode separation expanding
$\Psi(x^\mu)$ in terms of the $(n-1)$-dimensional spherical harmonics 
$Y_{\vec k}^l(x^i)$, which satisfy \cite{Vilenkin book}
\begin{equation}
\Delta^{(n-1)} Y_{\vec k}^l(x^i)= -l(l+n-2)Y_{\vec k}^l(x^i),  \label{10}
\end{equation}
where $l=0,1,2, ...$; $l\geq k_1\geq k_2\geq ...\geq k_{n-2}\geq 0; \vec k
=(k_1, ..., \pm k_{n-2})$. These generalized spherical harmonics form an
orthonormal basis of functions on the $(n-1)$-sphere,
\begin{equation}
\int \sqrt{\tilde g}dx^1...dx^{n-1} Y_{\vec k}^{l*}(x^i) 
Y_{\vec k^{\prime}}^{l^{\prime}}(x^i)= \delta^{ll^{\prime}} \delta_
{\vec k\vec k ^{\prime}},  
\label{11}
\end{equation}
and we may write,
\begin{equation}
\Psi(x^\mu)=\sum_{l=0}\sum_{\vec k} \Psi_{\vec k}^l(t)Y_{\vec k}^l(x^i).
\label{12}
\end{equation}

When $\Psi$ is a real field, the coefficients $\Psi _{\vec k}^l$ are not all
independent, for instance in three dimensions we simply have 
$\Psi_{\vec k}^{l*}=\Psi _{-\vec k}^l$. Now let us substitute (\ref{12}) into
(\ref{5}), use (\ref{10}) and note that $(n-2)^2/4+l(l+n-2)=(l+1+(n-4)/2)^2$. If
we also introduce a new index $k$ instead of $l$ by $k=l+1$, so that $k=1,2,...$
we obtain
\begin{equation}
S_m=\int dt\sum_{k=1}^\infty \sum_{\vec k}\left[ \dot\Psi_{\vec k}^{l*} 
\dot\Psi_{\vec k}^{l} -M_k^2 \Psi_{\vec k}^{l*} \Psi_{\vec k}^{l} + U
\Psi_{\vec  k}^{l*} \Psi_{\vec k}^{l} \right]  \label{13}
\end{equation}
where
\begin{equation}
M_k\equiv k+{\frac{n-4}{2}}.  \label{14}
\end{equation}

Note that the coefficients of (\ref{12}), $\Psi_{\vec k}^{l}(t)$ are just
functions of $t$ ($1$-dimensional fields), and for each set $(l,\vec k)$ we
may introduce two real functions $\phi_{\vec k}^{l}(t)$ and 
$\tilde\phi_{\vec k}^{l}(t)$ defined by
\begin{equation}
\Psi_{\vec k}^{l}\equiv {\frac{1}{\sqrt{2}}}\left(\phi_{\vec k}^{l}+ 
i\tilde\phi_{\vec k}^{l}\right),
\label{15}
\end{equation}
then the action (\ref{13}) becomes the sum of the actions of two independent
sets formed by an infinite collection of decoupled time dependent harmonic
oscillators
\begin{equation}
S_m={\frac{1}{2}}\int dt \sum_{k=1}^\infty \sum_{\vec k} 
\left[ \left( \dot\phi_{\vec k}^{l}\right)^2 -M_k^2 \left(\phi_{\vec
k}^{l}\right)^2 +U(t) \left(\phi_{\vec k}^{l}\right)^2 \right] + ...\, , 
\label{16} \end{equation}
where the dots stand for an identical action for the real $1$-dimensional
fields $\tilde \phi_{\vec k}^{l}(t)$.

We will consider, from now on, the action for the $1$-dimensional fields
$\phi _{\vec k}^l$ only. If our starting field $\Phi $ in (\ref{2}) is real
the results from this ``half'' action, i.e. the written term in (\ref{16}),
are enough, if $\Phi $ is complex we simply have to double the number of
degrees of freedom. Since $M_k$ depends on $k$ but not on $\vec k$, there is
no dependence in the action on the vector $\vec k$ and we can substitute
$\sum_{\vec k}$ by $\sum_{\vec k}1$ which gives the degeneracy of the mode
$k$. This is given by \cite{Garriga 1994},
\begin{equation}
\sum_{\vec k}1={\frac{(2k+n-4)(k+n-4)!}{(k-1)!(n-2)!}}.  \label{17}
\end{equation}
Note that for $n=2$, i.e. when the space section is a circle 
$\sum_{\vec k}1=2$; when $n=3$, which corresponds to the case of the ordinary
spherical harmonics $\sum_{\vec k}1=2k-1$ (or $2l+1$ in the usual notation); and
for $n=4$, which is the case of interest here, the space section of the
spacetime are 3-spheres and we have $\sum_{\vec k}1=k^2$.

The field equation for the $1$-dimensional fields $\phi_{\vec k}^{l}(t)$
are, from (\ref{16}),
\begin{equation}
\ddot\phi_{\vec k}^{l}+ M_k^2\phi_{\vec k}^{l}= U(t)\phi_{\vec k}^{l},
\label{18}
\end{equation}
which in accordance with our previous remarks will be solved perturbatibely,
being $U(t)$ the perturbative term. The solutions of the unperturbed
equation can be written as linear combinations of the normalized positive
and negative frequency modes, $f_k$ and $f_k^*$ respectively, where
\begin{equation}
f_k(t)={\frac{1}{\sqrt{2M_k}}}\exp (-iM_kt).  \label{19}
\end{equation}

\subsection{Closed time path effective action}

Let us now derive the semiclassical closed time path (CTP) effective action 
$\Gamma _{CTP}$ for the cosmological scalar factor due to the presence of the
quantum scalar field $\Phi $. The computation of the CTP effective action is
similar to the computation of the ordinary (in-out) effective action, except
that now we have to introduce two fields, the plus and minus fields $\phi
^{\pm }$, and use appropriate ``in'' boundary conditions. These two fields
basically represent the field $\phi $ propagating forward and backward in
time. This action was introduced by Schwinger \cite{Schwinger}
to derive expectation values rather than matrix elements as
in the ordinary effective action, and it has been used recently in connexion
with the back-reaction problem in semiclasical gravity \cite{CTP,CH94,CV94}.
Here we follow the notations and conventions of refs. \cite{CV97,CCV97}

Note that since we are considering the interaction of the scale factor $a$
with the quantum field $\phi$, in the CTP effective action we have now two
scalar fields $\phi^\pm$ and also two scale factors $a^\pm$. The kinetic
operators for our 1-dimensional fields $\phi_{\vec k}^l$ are given by
$A_k={\rm diag}(-\partial_t^2-M_k^2+U^+(t),\, \partial_t^2+M_k^2-U^-(t))$. The
propagators per each mode $k$, $G_k(t,t^\prime)$ are defined as usual by
$A_kG_k=\delta$, and are $2\times2$ matrices with components
$(G_k)_{\pm\,\pm}$.

To one loop order in the quantum fields $\phi^\pm$ and at three level in the
classical fields $a^\pm$ the CTP effective action for $a^\pm$ may be
written as,
\begin{equation}
\Gamma_{CTP}[a^\pm]=S_g[a^+]-S_g[a^-]+S_m^{cl}[a^+]-S_m^{cl}[a^-]- 
{\frac{i}{2}} \sum_{k=1}\sum_{\vec k}^\infty{\rm Tr} (\ln G_k),  
\label{20}
\end{equation}
where $S_g$ is the pure gravitational action, $S_m^{cl}$ is the action of
classical matter which in our case will include the cosmological constant
term only, and $G_k$ is the propagator for the mode $k$ which solves (\ref
{18}). In principle the $\Gamma _{CTP}$ depends on the expectation value in
the quantum state of interest, the ``in'' vacuum here, of both $a^{\pm }$
(the classical field) and of $\phi ^{\pm }$. To get the previous expression
we have substituted the solution of the dynamical equation for the
expectation value of the scalar field which is $\langle 0,in|\phi
|0,in\rangle =0$, so that there is no dependence on the expectation values
of $\phi ^{\pm }$ in the effective action.

Because of the interaction term $U(t)$ in (\ref{18}) the propagator $G_k$
cannot be found exactly and we treat it perturbatively. Thus we can write 
$G_k=G_k^0(1-UG_k^0+UG_k^0UG_k^0+\dots)$ where the unperturbed propagator is 
$(G_k^0)^{-1}={\rm diag}(-\partial_t^2-M_k^2,\, \partial_t^2 +M_k^2)$. This
unperturbed propagator has four components $(G_k^0)_{+\,+}=\Delta_{kF}$, 
$(G_k^0)_{-\,-}=-\Delta_{kD}$, $(G_k^0)_{+\,-}=-\Delta_{k}^+$ and 
$(G_k^0)_{-\,+}=\Delta_k^-$, where $\Delta_{kF}$, $\Delta_{kD}$ and 
$\Delta_k^{\pm}$ are the Feynman, Dyson and Wightman propagators for the mode 
$k$. This is a consequence of the boundary conditions which guarantee that
our quantum state is the ``in" vacuum $|0,in\rangle$. These propagators are
defined with the usual $i\epsilon$ prescription by
\begin{eqnarray}
\Delta _{kF}(t-t^{\prime }) &=&{\frac 1{2\pi }}\int_{-\infty }^\infty 
{\frac{\exp (-i\omega (t-t^{\prime }))}{\omega ^2-(M_k^2-i\epsilon )}} d\omega
\label{feynman} \\
&=&-i\left[ f_k(t)f_k^{\star }(t^{\prime })\theta (t-t^{\prime })+
f_k^{\star}(t)f_k(t^{\prime })\theta (t^{\prime }-t)\right] ,  \nonumber
\end{eqnarray}
\begin{eqnarray}
\Delta _{kD}(t-t^{\prime }) &=&{\frac 1{2\pi }}\int_{-\infty }^\infty 
{\frac{\exp (-i\omega (t-t^{\prime }))}{\omega ^2-(M_k^2+i\epsilon )}}d\omega
\label{dyson} \\
&=&i\left[ f_k^{\star }(t)f_k(t^{\prime })\theta (t-t^{\prime
})+f_k(t)f_k^{\star }(t^{\prime })\theta (t^{\prime }-t)\right] ,  \nonumber
\end{eqnarray}
\begin{equation}
\Delta_k^+(t-t^\prime)=if_k^\star(t)f_k(t^\prime),\ \ \ \ \
\Delta_k^-(t-t^\prime)=-if_k(t)f_k^\star(t^\prime).  \label{21c}
\end{equation}

The trace term in the effective action (\ref{20}) will now be expanded up to
order $\nu^2$. The linear terms in $\nu$ are tadpoles which are zero in
dimensional regularization. Thus we can write the effective action as
\begin{equation}
\Gamma_{CTP}[a^\pm]\simeq S_g[a^+]-S_g[a^-]+S_m^{cl}[a^+]-S_m^{cl}[a^-] +T^+
+T^- +T,  \label{22}
\end{equation}
where
\begin{equation}
T^{\pm }=-{\frac i4}\sum_{k=1}^\infty \sum_{\vec k}{\rm Tr}\left(
U_{\pm}(G_k^0)_{\pm \,\pm }U_{\pm }(G_k^0)_{\pm \,\pm }\right) ,\ \ \ \ 
T=\frac {i}{2}\sum_{k=1}^\infty \sum_{\vec k}{\rm Tr}\left(
U_{+}(G_k^0)_{+\,-}U_{-}(G_k^0)_{-\,+}\right) .  \label{23}
\end{equation}

The pure gravitational part of the action, $S_g$, includes the
Einstein-Hilbert action and a quadratic counterterm which is needed for
regularization of the divergences of (\ref{23}),
\begin{equation}
S_g={\frac{1}{l_P^2}} \int d^n x\sqrt{-g} R + {\frac{\nu^2 \mu_c^{n-4}}
{32\pi^2 (n-4)}} \int d^n x\sqrt{-g} R^2,  \label{24}
\end{equation}
where $\mu_c$ is an arbitrary mass scale which gives the correct dimension
to the counterterm, and $l_P^2=16\pi G$, the square of the Planck length. To
regularize the divergencies in $T^\pm$ we need to expand the action 
(\ref{24}) in powers of $n-4$. Using our metric (\ref{1}), we can perform the
space integration in (\ref{24}) which leads to the volume to the $(n-1)$-sphere.
Expanding now in powers of $n-4$, and recalling that the volume of the
three-sphere is $2\pi^2$ we may write $S_g=S_g^{EH}+S_g^{div}$, where the
first term stands for the Einstein-Hilbert action in four dimensions and the
second term is the first order correction in this expansion,
\begin{eqnarray}
S_g[a]&=&{\frac{2\pi^2}{l_P^2}} \int dt\, 6a^2\left( {\frac{\ddot a}{a}}
+1\right),  \label{25a} \\
S_g^{div}[a,\mu_c]&=&{\frac{1}{16}}\left\{ {\frac{1}{n-4}}\int dt U_1^2(t)+
\int dt \left[ U_1^2(t)\ln (a\mu_c) + 2 U_1(t)U_2(t) \right] \right\}.
\label{25b}
\end{eqnarray}
Here $U_1(t)$ and $U_2(t)$ are defined by the expansion of $U$ in powers of
$n-4$. That is, from (\ref{8}) and (\ref{3}) we can write $U(t)=
U_1(t)+(n-4)\, U_2(t)$, where
\begin{equation}
U_1=-6\nu\left({\frac{\ddot a}{a}}+1\right),\ \ \ \ U_2=-\nu\left(2 
{\frac{\ddot a}{a}}+3{\frac{\dot a^2}{a^2}}+5\right).  
\label{9}
\end{equation}

The classical matter term $S_m^{cl}$ includes in our case the cosmological
constant $\Lambda^\star$ only. It can be understood as the term which gives
the effect of the inflaton field at the stationary point of the inflaton
potential \cite{Vilenkin 1982-1984}:
\begin{equation}
S_m^{cl}[a]=-2\pi^2 \int dt a^4\Lambda^\star .  \label{26}
\end{equation}

\subsection{Computation of $T$ an $T^\pm$}

Let us first compute $T$ in (\ref{23}), which may be written as
\begin{equation}
T=-{\frac i2}\sum_{k=1}^\infty \sum_{\vec k}\int dtdt^{\prime
}U^{+}(t)\Delta _k^{+}(t-t^{\prime })U^{-}(t^{\prime })\Delta
_k^{-}(t^{\prime }-t).  \label{27}
\end{equation}
Since this term will not diverge we can perform the computation directly in
$n=4$ dimensions. In this case $\sum_{\vec k}1=k^2$ and $M_k=k$, thus using
(\ref{21c}) and (\ref{19}) we have
\begin{eqnarray}
T &=&-{\frac i2}\int dtdt^{\prime }\sum_{k=1}^\infty
U^{+}(t)k^2f_k^{*2}(t)f_k^2(t^{\prime })U^{-}(t^{\prime })  \nonumber \\
&=&-\int dtdt^{\prime }U^{+}(t)D(t-t^{\prime })U^{-}(t^{\prime })-i\int
dtdt^{\prime }U^{+}(t)N(t-t^{\prime })U^{-}(t^{\prime }),  \label{28}
\end{eqnarray}
where we have introduced the kernels $D$ and $N$ as,
\begin{eqnarray}
D(t-t^{\prime})&\equiv&-{\frac{1}{8}}\sum_{k=1}^\infty \sin 2k(t-t^{\prime})
=-{\frac{1}{16}} {\rm PV}\left[ {\frac{\cos (t-t^{\prime})}{\sin
(t-t^{\prime})}} \right]  \label{29a} \\
N(t-t^{\prime})&\equiv&{\frac{1}{8}}\sum_{k=1}^\infty \cos 2k(t-t^{\prime})
={\frac{1}{16}}\left\{ \pi \left[ \sum_{n=\infty}^\infty \delta
(t-t^{\prime}-n\pi) \right] -1 \right\},  \label{29b}
\end{eqnarray}
and we have computed the corresponding series. The kernels $D$ and $N$ are
called dissipation and noise kernel, respectively, using the definitions of 
\cite{CCV97}. It is interesting to compare with Ref. \cite{CV97,CCV97} where
a spatially flat universe was considered. Our results may be formaly
obtained from that reference if we change $vol \int_0^\infty dk$ there,
where $vol$ is the volume of the space section (assume for instance a finite
box), by $2\pi^2\sum_{k=1}^\infty$. In the spatially flat case the noise is
a simple delta function (white noise), whereas here we have a train of
deltas. Note also that we have, in practice, considered a real scalar field
only since we considered only half of the action, i.e. the written part of
(\ref{16}). Thus for the complex scalar field we need to
multiply these kernels by two, i.e. the dissipation kernel is $2D$ and the
noise kernel is $2N$. Note also that the definition of the dissipation
kernel here and in Ref. \cite{CH94} differ by a sign.

Let us now perform the more complicated calculation of $T^\pm$. Since these
integrals diverge in $n=4$ we work here in arbitrary $n$ (dimensional
regularization). From (\ref{23}) and the symmetries of $\Delta_{kF}$ and
$\Delta_{kD}$ we have
\begin{equation}
T^\pm=-{\frac{i}{4}}\int dt dt^{\prime}U^\pm(t) \Delta_{F/D}^2(t-t^{\prime})
U^\pm (t^{\prime}),  \label{30}
\end{equation}
where we have introduced,
\begin{equation}
\Delta _{F/D}^2(t-t^{\prime })\equiv \sum_{k=1}^\infty 
\left( \sum_{\vec k}1\right) \Delta _{kF/D}^2(t-t^{\prime })=\int_{-\infty
}^\infty {\frac{d\omega }{2\pi }}e^{-i\omega (t-t^{\prime })}I(\omega),
\label{31} \end{equation}
where $I(\omega)$ is defined after having made an integral in $\omega$ with
appropriate contour, recall the definitions (\ref{feynman}) and 
(\ref{dyson}). After using (\ref{17}) and the definition (\ref{14}) of $M_k$,
$I(\omega ) $ is given by
\begin{eqnarray}
I(\omega)&=&\pm {\frac{i}{2(n-2)!}}\sum_{k=1}^\infty {\frac{(k+n-4)!}
{(k-1)!\left[( k+(n-4)/2)^2-(\omega/2)^2 \pm i0^+\right]}}  \nonumber \\
&\equiv& \pm {\frac{i}{2(n-2)!}}\sum_{k=1}^\infty a_k (\omega),  \label{32}
\end{eqnarray}
where we have introduced the coefficients $a_k$ in the last series
expression. In Appendix A we prove that this series diverges like
$1/(n-4)$, and thus we can regularize it using (\ref{25b}). Furthermore its
imaginary part is finite and leads to the noise kernel $N$ defined above.

Thus according to (\ref{33}), (\ref{35}) and (\ref{39bis}) from the Appendix
we can write (\ref{31}) as
\begin{equation}
\Delta _{F/D}^2(t-t^{\prime })=\mp \left( \frac i4\right) \left[ 
{\frac{\delta (t-t^{\prime })}{n-4}}-{\frac 1{2\pi ^2}}K^{\pm }(t-t^{\prime
})\right] ,  \label{40}
\end{equation}
where, we have defined
\begin{equation}
K^{\pm }(t-t^{\prime })\equiv 16\pi ^2\left[ A(t-t^{\prime })\pm
iN(t-t^{\prime })\right] .  \label{41}
\end{equation}
Here $A(t-t^{\prime })$ is a finite kernel which will be discussed below. We
can now substitute (\ref{40}) into (\ref{30}) and use the expansion of $U(t)$
in powers of $n-4$ given in (\ref{9}) to get
\begin{equation}
T^\pm=\mp\left[ {\frac{1}{n-4}}\int dt (U_1^\pm)^2+ 2\int dt U_1^\pm U_2^\pm
-{\frac{1}{2\pi^2}}\int dt dt^{\prime}U_1^\pm (t) K^\pm (t-t^{\prime})
U_1^\pm (t^{\prime}) \right].  \label{42}
\end{equation}

\subsection{The regularized CTP effective action}

We are now in the position to compute the regularized semiclassical CTP
effective action. Let us substitute in (\ref{22}) the actions (\ref{25a}),
(\ref{25b}) and (\ref{26}), and the results (\ref{28}) for $T$ and (\ref{42})
for $T^\pm$. It is clear that the divergent term in (\ref{42}), i.e. the
term proportional to $1/(n-4)$, will be cancelled by the divergent
counterterm in (\ref{25b}). Also the terms $\int dt U_1 U_2$ in these
equations will cancel. Thus, we finally get the regularized semiclassical
action
\begin{equation}
\Gamma_{CTP}[a^\pm]= S_{g,m}^R[a^+]-S_{g,m}^R[a^-]+ S_{IF}^R[a^\pm],
\label{43a}
\end{equation}
where the regularized gravitational and classical matter actions are,
\begin{equation}
S_{g,m}^R[a]= {\frac{2\pi^2}{l_P^2}}\int dt\, 6a^2\left( 
{\frac{\ddot a}{a}}+1\right) -2\pi^2 \int dt\, a^4 \Lambda^\star
+{\frac{1}{16}}\int dt\, U_1^2(t) \ln (a\mu_c).  \label{43b}
\end{equation}

To write the remaining part, $S_{IF}^R$, we note that the kernels $A$ and $N$
in (\ref{41}), satisfy the symmetries $A(t-t^{\prime})=A(t^{\prime}-t)$ and
$N(t-t^{\prime})=N(t^{\prime}-t)$. Taking into account also that
$D(t-t^{\prime})=-D(t^{\prime}-t)$ we obtain
\begin{equation}
S_{IF}^R[a^\pm]={\frac{1}{2}} \int dt dt^{\prime}\, \Delta U(t)
H(t-t^{\prime}) \{ U(t^{\prime})\} +{\frac{i}{2}} \int dt dt^{\prime}\,
\Delta U(t) N(t-t^{\prime}) \Delta U(t^{\prime}),  \label{43c}
\end{equation}
where we have defined
\begin{equation}
H(t-t^{\prime})= A(t-t^{\prime};\mu_c)-D(t-t^{\prime}),  \label{44a}
\end{equation}
\begin{equation}
\Delta U= U^+-U^-,\ \ \ \ \ \{ U\} = U^+ +U^-.  \label{44b}
\end{equation}
In (\ref{44a}) we have explicitly written that the kernel $A$ depends on the
renormalization parameter $\mu _c$. We note that this effective action has
an imaginary part which involves the noise kernel $N$. However, because of
the quadratic dependence of this term in $\Delta U$ it will not contribute
to the field equations if we derive such equations from $\delta \Gamma
_{CTP}/\delta a^{+}|_{a^{\pm }=a}=0$. This, in fact, gives the dynamical
equations for expectation values of the field $a(t)$.

However, we recall that we are dealing with the interaccion of a ``system'',
our classical (one dimensional) field $a(t),$ with an ``environment'' formed
by the degrees of freedom of the quantum system and that we have integrated
out the degrees of freedom of the environment (note that in the effective
action we have substituted the solutions of the field equations for the
expectation value of the quantum field). In this case the regularized action 
$S_{IF}^R$ can be understood as the influence action of the
system-environment interaction, which describes the effect of the
environment on the system of interest \cite{Feynman,Hu89}. The imaginary
part of the influence action is known \cite{CH94,hargell,Gleiser and Ramos
1994,CH95,HuSin,CV96,eft} to give the effect of a stochastic force on the
system, and we can introduce an improved semiclassical effective action,
\begin{equation}
S_{eff}[a^{\pm };\xi ]=S_{g,m}^R[a^{+}]-S_{g,m}^R[a^{-}]+{\frac 12}\int
dtdt^{\prime }\,\Delta U(t)H(t-t^{\prime })\{U(t^{\prime })\}+\int dt\xi
(t)\Delta U(t),  \label{45a}
\end{equation}
where $\xi(t)$ is a Gaussian stochastic field defined by the following
statistical averages
\begin{equation}
\langle \xi(t) \rangle =0,\ \ \ \ \langle \xi(t) \xi(t^{\prime}) \rangle
=N(t-t^{\prime}).  \label{45b}
\end{equation}
The kernel $H$ in the effective action gives a non local effect
(due to particle creation), whereas the source $\xi$ gives the
reaction of the environment into the system in terms of a stochastic force.

The formal derivation of the last term of (\ref{45a}) can be seen as follows.
The Feynman and Vernon influence functional \cite{Feynman} of the
system-environment interaction is defined from the influence action $S_{IF}$ by
$F_{IF}=\exp (iS_{IF})$. Note now that by 
using a simple path integral Gaussian identity, the imaginary part of
(\ref{43c}) can be formally recovered in $F_{IF}$ with the following functional
Fourier transform
$F_{IF}=\int D\xi P[\xi]\exp\left\{ i[{\rm Re}(S_{IF})+\int dt\xi(t)\Delta
U(t)]\right\} $, where 
$$P[\xi]=\frac {\exp [-{1\over2}\int dt dt'\xi(t)N^{-1}(t-t')\xi(t')]}
{\int D\xi \exp \left[-{1\over2}
\int dt dt'\xi(t)N^{-1}(t-t')\xi(t')\right]},$$
can be interpreted as a Gaussian probability distribution for the
field $\xi$. That is, the influence funcional may be seen as the statistical
average of $\xi$ dependent influence functionals constructed with the
``effective'' influence action ${\rm Re}(S_{IF})+\int dt\xi(t)\Delta U(t)$. The
physical interpretation of this result, namely, that
the semiclassical equations are now the stochastic
equations derived from such effective action
may be seen, for instance, in Ref. \cite{hargell}.

\subsection{Stochastic semiclassical back-reaction equation}

The dynamical equation for the scale factor $a(t)$ can now be found from the
effective action (\ref{45a}) in the usual way, that is by functional
derivation with respect to $a^{+}(t)$ and then equating 
$a^{+}=a^{-}\equiv a$. These equations include the back-reaction of the quantum
field on the scale factor. It is convenient to use a rescaled scale factor $b$
and cosmological constant $\Lambda $ defined by
\begin{equation}
b(t)\equiv {\frac{\sqrt{24}\pi}{l_P}} a(t),\ \ \ \ \ \ \Lambda\equiv 
{\frac{l_P^4}{12\pi^2}} \Lambda^\star.  
\label{46}
\end{equation}

The regularized action $S_{g,m}^R$ becomes after one integration by parts,
\begin{equation}
S_{g,m}^R[b]=-{\frac{1}{2}}\int dt \left[ \dot b^2 -b^2 +{\frac{1}{12}}
\Lambda b^4 -{\frac{9}{2}}\nu^2 \left( {\frac{\ddot b}{b}}+1\right)^2 \ln (b 
\bar \mu), \right]  \label{47}
\end{equation}
where we have also rescaled the renormalization parameter $\bar \mu$. The
remaining term in (\ref{45a}) does not change with this rescaling except
that now $U(t)$ should be written in terms of $b$, thus according to 
(\ref{9}) we have
\begin{equation}
U(t)=-6\nu\left({\frac{\ddot b}{b}}+1\right).  \label{48}
\end{equation}

The dynamical equation for $b(t)$ is:
\begin{equation}
\left.{\frac{\delta S_{eff}[b^\pm;\xi] }{\delta b^+}}\right|_{b^\pm=b} =0.
\label{49}
\end{equation}
This equation improves the semiclassical equation by taking into account the
fluctuations of the stress-energy tensor of the quantum field \cite
{fordkuo,hunich,Martin and Verdaguer 1998}. When averaged over $\xi$ the equation
leads to the usual semiclassical equation for the expectation value of $b(t)$.

Now this equation leads to the typical non physical runaway solutions due to
the higher order time derivatives involved in the quantum correction terms.
To avoid such spurious solutions we use the method of order reduction \cite
{Flanagan and Wald 1996} into the equations (\ref{49}). In this method one
asumes that equation (\ref{49}) are perturbative equations in which the
perturbations are the quantum corrections. To leading order the equation
reduces to the classical equation
\begin{equation}
\ddot b+b\left( 1-{\frac{1}{6}}\Lambda b^2\right)= O(\nu).  \label{50}
\end{equation}
The terms with $\ddot b$ or with higher time derivatives in the quantum
corrections of the equation (\ref{49}) are then substituted using
recurrently the classical equation (\ref{50}). In this form the solutions to
the semiclassical equations are also perturbations of the classical
solutions. Thus by functional derivation of (\ref{45a}), using (\ref{47}),
we can write the stochastic semiclassical back-reaction equation (\ref{49})
as
\begin{equation}
\dot p=-V^{\prime }\left( b\right) -\delta V^{\prime }(b)+ F(b,p,t)
+J(\xi,b,p),  \label{one}
\end{equation}
where a prime means a derivative with respect to $b$, and we have introduced 
$p\equiv\dot b$. The classical potential $V(b)$ is
\begin{equation}
V\left( b\right) = \frac 12 b^2- \frac \Lambda {24} b^4,  \label{two}
\end{equation}
and its local quantum correction is
\begin{equation}
\delta V (b) = - \frac{3\nu ^2\Lambda }4 \left[ \frac 12 b^2- 
\frac \Lambda {48} b^4-p^2\ln (b\bar \mu) \right],  
\label{three}
\end{equation}
where we have implemented order reduction in this term. On the other hand
the term $F(b,p,t)$ involves nonlocal contributions and may be written as,
\begin{equation}
F(b,p,t)=-\frac{\partial U}{\partial b}I-\frac{d^2}{dt^2}\left( 
\frac{\partial U}{\partial \ddot b}I\right) =
6\nu \left( \frac{d^2}{dt^2}\frac 1b-\frac{\ddot b}{b^2}\right) I,  
\label{51}
\end{equation}
where $I(b,p,t)$ is defined by
\begin{equation}
I(b,p,t)\equiv \int_{-\infty }^\infty dt^{\prime }H(t-t^{\prime
})U(t^{\prime }).  \label{52}
\end{equation}
After order reduction, $U\left( t^{\prime }\right) $ must be evaluated on
the classical orbit with Cauchy data $b\left( t\right) =b$, $p\left(
t\right) =p$, whereby it reduces to $U=-\Lambda \nu b^2$.
The function $J$ is the noise given by
\[
J\left( \xi ,b\right) =6\nu \left[ \frac{d^2}{dt^2}\left( 
\frac \xi b \right) -\frac{\ddot b\xi }{b^2}\right] 
\]
and, after order reduction, by
\begin{equation}
J(\xi,b,p) =6\nu \left[ \frac{\ddot \xi }b-\frac{2\dot \xi p}{b^2} +
\frac{2\xi V^{\prime }\left( b\right) }{b^2}+\frac{2\xi p^2}{b^3}\right],
\label{six}
\end{equation}
with $\xi (t)$ defined in (\ref{45b}) in terms of the noise kernel.

\subsection{Approximate kernels $N$ and $H$}

To simplify the nonlocal term $F(b,p,t)$ and the noise $J(\xi,b,p)$
we will approximate the kernel $H$ and the noise kernel $N$ 
keeping only the first delta function, i.e. $n=0$, in the train
of deltas which define the noise kernel $N$. This amounts to take the
continuous limit in $k$ in the definition (\ref{29b}) of $N$. In fact, we
take the sum in $k$ as an integral and we get
\begin{equation}
N(u)=\int_0^\infty dk \cos 2ku ={\frac{\pi}{16}}\delta(u).  \label{53a}
\end{equation}

This is equivalent to assume that the spacetime spatial sections are flat
and of volume $2\pi^2$, see Ref.\cite{CV97}. Similarly the dissipation
kernel $D$ defined in (\ref{29a}) becomes
\begin{equation}
D(u)=-{\frac{1}{8}}\int_0^\infty dk \sin 2ku= -{\frac{1}{16}}{\rm PV} \left( 
{\frac{1}{u}}\right).  \label{53b}
\end{equation}

The same approximation may be used to compute the kernel $A$ defined in (\ref
{32})-(\ref{41}). The computation of this kernel can be read directly from
(\ref{39}), see also Ref. \cite{CV97}
\begin{equation}
A(u)=-{\frac{1}{8}} \int_{-\infty}^\infty {\frac{d\omega}{2\pi}} e^{-i\omega
u} \ln {\frac{|\omega|}{|\mu_c|}}= {\frac{1}{16}}{\rm Pf}\left( 
{\frac{1}{|u|}}\right) +{\frac{1}{8}}(\gamma +\ln \mu_c)\delta(u),  
\label{53c}
\end{equation}
where $\gamma$ is Euler's number and Pf means the Hadamard principal
function whose meaning will be recalled shortly. To perform this last
Fourier tranform we write $\ln |\omega|=\lim_{\epsilon\rightarrow 0^+}[\exp
(-\epsilon|\omega|) \ln |\omega|]$, use the integrals $\int_0^\infty
d\omega\ln \omega \cos (\omega u)\exp (-\epsilon\omega)$ and $\int_0^\infty
d\omega \cos (\omega u)\exp (-\epsilon\omega)$ which can be found in \cite
{Gradshteyn 1979}, and take into account that 
$[2x\tan^{-1}(u/\epsilon)+\epsilon\ln(u^2+\epsilon^2)]/ 
(u^2+\epsilon^2)= d[\ln(u^2+\epsilon^2)\tan^{-1}(u/\epsilon)]/du$. When
$\epsilon\rightarrow 0^+ $ the last expression gives a representation of
$\pi{\rm Pf}(1/|u|)$. Finally, using (\ref{53b}) and (\ref{53c}) the kernel of
interest $H(u)=A(u)-D(u)$ can be written as,
\begin{equation}
H(u)=\frac {1}{8} {\rm Pf}\left[ \frac {\theta (u)}{u}\right] + 
\frac {\gamma+\ln\mu_c}{8} \delta (u).  \label{53d}
\end{equation}

The distribution ${\rm Pf}(\theta(u)/u)$ should be understood as follows.
Let $f(u)$ be an arbitrary tempered function, then
\begin{equation}
\int_{-\infty}^\infty du {\rm Pf}\left[ \frac {\theta (u)}{u}\right] f(u)=
\lim_{\epsilon\rightarrow 0^+}\left( \int_\epsilon^\infty du \frac {f(u)}{u}
+f(0)\ln\epsilon\right).  \label{53e}
\end{equation}

The approximation of substituting the exact kernels by their flat space 
counterparts is clearly justified when the radius of the universe is large,
which is when the semiclassical approximation works best.
Once the local approximation for the noise kernel follows, the corresponding
expression for $D$ can be obtained by demanding that their Fourier transforms be
related by the same fluctuation-dissipation relation as in the exact formula.

\section{The Fokker-Planck equation}
\label{section3}

Now we want to determine the probability that a universe starting at the
potential well goes over the potential barrier into the inflationary stage.
In statistical mechanics this problem is known as Kramers' problem. To
describe such process we have the semiclassical back-reaction equation (\ref
{one}), which is a stochastic differential equation (a Langevin type of
equation). As it is well known \cite{Risken 1984} to study this problem it is
better to construct a Fokker-Planck equation, which is an ordinary
differential equation for a distribution function. Thus, the first step will
be to derive the Fokker-Planck equation corresponding to the stochastic
equation (\ref{one}). The key features of this stochastic equation are: a
potential given by the local potentials (\ref{two}) and (\ref{three}), a non
local term given by the function $F$ and a noise term $J$. The classical
part of the potential has a local minimum at $b=0$ then reaches a maximum
and decreases continuously after that. The inflationary stage corresponds to
the classical values of $b$ beyond this potential barrier. If we start near
$b=0$ the noise term will take the scale factor eventually over the barrier,
but if we want to compute the escape probability we need to consider both
noise and non locality.

It should do no harm if we disregard the local quantum correction to the
potential, $\delta V(b)$, the reason is the following. This term is a
consequence of renormalization, but in semiclassical gravity there is a two
parameter ambiguity in terms which are quadratic in the curvature in the
gravitational part of the action. This ambiguity is seen here only in the
parameter $\bar \mu $ because we have simply ignored the other possible
parameter which was not essential in the renormalization scheme. Furthermore
we should not trust the semiclassical results too close to $b=0$, since the
semiclassical theory should break down here. Thus the possible divergence at 
$b=0$ may be disregarded and we should think of this renormalized term as
just a small correction to the classical potential, as it is indeed for
all radii of the universe unless $b\ll 1$. Thus the classical
potential $V(b)$ should contain the main qualitative features of the local
renormalized potential. 

To construct the Fokker-Planck equation let us introduce the distribution
function
\begin{equation}
f\left( b,p,t\right) =\left\langle \delta \left( b\left( t\right) -b\right)
\delta \left( p\left( t\right) -p\right) \right\rangle,  \label{seven}
\end{equation}
where $b(t)$ and $p(t)$ are solutions of equation (\ref{one}) for a given
realization of $\xi (t)$, $b$ and $p$ are points in the phase space, and the
average is taken both with respect to the initial conditions and to the
history of the noise as follows. One starts by considering the ensemble of
systems in phase space obeying equation (\ref{one}) for a given realization
of $\xi (t)$ and different initial conditions. This ensemble is described by
the density $\rho (b,p,t)=\langle \delta (b(t)-b)\delta (p(t)-p)\rangle $,
where the average is over initial conditions. Next one defines the
probability density $f(b,p,t)$ as the statistical average over the
realizations of $\xi (t)$, that is $f(b,p,t)=\langle \rho (b,p,t)\rangle
_\xi $

The next manipulations are standard \cite{Sancho}, we take the time
derivative of $f$, 
\[
\partial _tf=\langle \dot b\left( t\right) \partial _{b(t)}\delta \left(
b\left( t\right) -b\right) \delta \left( p\left( t\right) -p\right) +\delta
\left( b\left( t\right) -b\right) \dot p\left( t\right) \partial
_{p(t)}\delta \left( p\left( t\right) -p\right) \rangle , 
\]
and note that 
$
\partial _{b(t)}\delta \left( b\left( t\right) -b\right) =-\partial _b\delta
\left( b\left( t\right) -b\right), 
$
and that 
$
\left\langle p\left( t\right) \delta \left( b\left( t\right) -b\right)
\delta \left( p\left( t\right) -p\right) \right\rangle =pf\left(
b,p,t\right). 
$

Performing similar manipulations for the other terms and using the equations
of motion (\ref{one}) we find
\begin{equation}
\frac {\partial f}{\partial t}= \{ H,f\} -\frac {\partial}{\partial p}
[F(b,p,t)f] -\frac {\partial}{\partial p} \Phi,  \label{eight}
\end{equation}
where we have defined
\begin{equation}
H(b,p)= \frac{1}{2} p^2+ V(b),  \label{54a}
\end{equation}
thus disregarding the potential $\delta V(b)$ in (\ref{one}), the curly
brackets are Poisson brackets, i.e. 
\[
\{H,f\}=-p(\partial f/\partial b)+V^{\prime }(b)(\partial f/\partial p), 
\]
and
\begin{equation}
\Phi =\left\langle J\left( \xi ,b,p\right) \delta \left( b\left( t\right)
-b\right) \delta \left( p\left( t\right) -p\right) \right\rangle.
\label{54b}
\end{equation}

Equation (\ref{eight}) is not yet a Fokker-Planck equation, to make it one
we need to write $\Phi$ in terms of the distribution function $f$. This term
will be called the diffusion term since it depends on the stochastic field 
$\xi(t)$.

From (\ref{54b}) and (\ref{six}) we may write
\begin{equation}
\Phi=6\nu \left[ \frac {C_2}{b}-\frac {2C_1p}{b^2} +\left(\frac {2V^\prime}
{b^2}+\frac {2p^2}{b^3}\right) C_0\right]  
\label{Phi}
\end{equation}
where
\begin{equation}
C_n=\left\langle \left(\frac {d^n}{dt^n}\xi(t)\right)
\delta(b(t)-b)\delta(p(t)-p)\right\rangle, \label{C_i}
\end{equation}
for $n=0,1,2$. To manipulate the difussion term of (\ref{eight}) we will
make use of the functional formula for Gaussian averages \cite{Novikov},
\begin{equation}
\left\langle \xi (t)R\left[ b\left( t\right) ,p\left( t\right) \right]
\right\rangle =\int dt^{\prime }\,N\left( t-t^{\prime }\right) 
\left\langle\frac \delta {\delta \xi \left( t^{\prime }\right) }
R\left[ b\left( t\right) ,p\left( t\right) \right] \right\rangle , 
\label{novikov} \end{equation}
where $R$ is an arbitrary functional of $\xi (t)$. Under the approximation
(\ref{53a}) for the noise kernel
\begin{equation}
C_0=\frac \pi {16}\left. \left\langle\frac \delta {\delta \xi \left(
t^{\prime}\right) } \delta \left( b\left( t\right) -b\right) \delta \left(
p\left( t\right) -p\right) \right\rangle\right|_{t^{\prime }\rightarrow t} =
-\frac{\pi \nu \Lambda }8b\frac \partial {\partial p}f\left( b,p,t\right),
\label{C_0}
\end{equation}
where we have used (\ref{twelve}) in the last step. The expressions for $C_1$
and $C_2$ are similarly obtained, first one uses the time translation
invariance of the noise kernel to perform integration by parts, then the
problem reduces to taking time derivatives of (\ref{C_0}). The results are
(see Appendix B for details) 
\begin{eqnarray}
C_1&=&(\pi \nu \Lambda /8)(b\partial _bf-p\partial _pf),  \label{C_1}
\\
C_2&=&(\pi \nu \Lambda /8)(2p\partial _bf+V^{\prime }\partial _pf+bV^{\prime
\prime }\partial _pf).  \label{C_2}
\end{eqnarray}

Finally, after substitution in (\ref{Phi}) and using the equation of motion
to lowest order we have
\begin{equation}
\Phi =-\frac{\pi \nu ^2\Lambda ^2}4b^2\frac{\partial f}{\partial p},
\label{Phifinal}
\end{equation}
which by (\ref{eight}) leads to the final form of the Fokker-Planck equation
\begin{equation}
\frac{\partial f}{\partial t}=\{ H,f\} -\frac {\partial}{\partial p}
[F(b,p,t)f] +\frac{\pi \nu ^2\Lambda ^2}{4} b^2\frac{\partial ^2f}{\partial
p^2}  \label{FokkerPlanck}
\end{equation}

We also notice that in the absense of a cosmological constant, we get no
diffusion. This makes sense, because in that case the classical trajectories
describe a radiation filled universe. Such universe would have no scalar
curvature, and so it should be insensitive to the value of $\nu $ as well.

\subsection{Averaging over angles}

We want to compute the probability that a classical universe
trapped in the potential well of $V(b)$ goes over the potential barrier as a
consequence of the noise and non locality produced by the interacction with
the quantum field, and end up in the de Sitter phase. A universe that
crosses this potential barrier will reach the de Sitter phase with some
energy which one would expect will correspond to the energy of the quantum
particles created in the previous stage. Note that this differs from the
quantum tunneling from nothing approach in which the universe gets to the de
Sitter stage tunneling from the potential minimum $b=0$ with zero energy. In
practice, this difference will not be so important because as the universe
inflates any amount of energy density will be diluted away.

For this computation we will follow closely the solution of Kramers' problem 
\cite{Kramers} reviewed in Appendix C. The three key features of such
computation are: first, the introduction of action-angle canonical variables
($J,\theta $); second, the asumption that $f$ depends on $J$ only, i.e.
$f(J) $; and, third, the use of the averaged Fokker-Planck equation over the
angle variable $\theta $. Of course, the Fokker-Planck equation in Kramers'
problem, (\ref{a1}), is much simpler than our equation (\ref{FokkerPlanck})
due to the non local character of the latter; thus we need to take care of
this problem, and it is quite remarkable that a relatively simple solution
can be found.

Thus, let us consider equation (\ref{FokkerPlanck}), introduce ($J, \theta$)
and assume that $f(J)$, in the Appendix we have seen that the dissipation
term which involves $\partial^2 f/\partial p^2$ can be written in terms of
derivatives with respect to $J$, see equation (\ref{a4}). Since we now have
$\{H,f\}=0$ we can write
\begin{equation}
\frac{\partial f}{\partial t}=\frac{\pi \nu ^2\Lambda ^2}4b^2\left( 
\frac 1\Omega \frac{\partial f}{\partial J}+\frac{p^2}\Omega 
\frac \partial {\partial J}\left( \frac 1\Omega 
\frac{\partial f}{\partial J}\right) \right) -\left[ 
\frac \partial {\partial p}F(b,p,t)\right] f-\frac{pF}\Omega 
\frac{\partial f}{\partial J}.  \label{b1} 
\end{equation}

Next we take the average of (\ref{b1}) with respect to the angle $\theta$.
The averaged equation involves the two pairs of integrals $\int d\theta b^2
p^2$, $\int d\theta b^2$, and $\int d\theta pF$, 
$\int d\theta \partial_p F$.
The components of each pair are related by a derivative with respect to 
$J$. In fact, let us introduce
\begin{equation}
{\bf D}(J)=\frac{1}{2\pi\Omega}\int_0^{2\pi}d\theta b^2 p^2,  \label{b2}
\end{equation}
changing the integration variable to $b$, see Appendix C, this integral may be
written as $\Omega\oint db b^2 p$, and using that $\partial_J p|_b=\Omega/p$
we have
\begin{equation}
\frac{d{\bf D}}{dJ}=\frac{1}{2\pi}\int d\theta b^2.  \label{b3}
\end{equation}

Similarly, let us introduce
\begin{equation}
S(J)=\frac{1}{2\pi\Omega}\int_0^{2\pi} d\theta p F(b,p,t),  \label{b4}
\end{equation}
again by a change of integration variable this integral may be written as
$\Omega\oint db F$ and by derivation with respect to $J$ we get
\begin{equation}
\frac{dS}{dJ}=\frac{1}{2\pi}\int_0^{2\pi} d\theta \frac{\partial}{\partial p}
F(b,p,t).  \label{b5}
\end{equation}

Finally, the average of the Fokker-Planck equation (\ref{b1}) becomes,
\begin{equation}
\frac{\partial f}{\partial t}=\frac{\pi\nu^2\Lambda^2}{4} 
\frac{\partial}{\partial J}\left[\frac{{\bf D}(J)}{\Omega} 
\frac{\partial f}{\partial J} \right] -\frac{\partial}{\partial J}(Sf). 
\label{b6} 
\end{equation}
This equation may be written as a continuity equation $\partial _tf+\partial
_JK=0$, where the probability flux $K$ may be identified directly from (\ref
{b6}). We see that, as in Kramers' problem, stationary solutions with
positive flux $K_0$ should satisfy
\begin{equation}
\frac{\pi\nu^2\Lambda^2}{4}\frac{{\bf D}(J)}{\Omega} 
\frac{\partial f}{\partial J}-S(J)f=-K_0.  \label{b7}
\end{equation}

\subsection{The non local contribution $S(J)$}

We need to handle now the term $S(J)$, defined in (\ref{b4}). The
problem here lies in the non local term $F(b,p,t)$ defined in 
(\ref{51})-(\ref{52}), with $U(t)$ given by (\ref{48}). Since this term 
gives a quantum correction to a classical equation we will adopt the order
reduction prescription. Thus let us assume that $b(t^{\prime })$ and
$p(t^{\prime })$ in the integral which defines $F$ are solutions to the
classical equations of motion with Cauchy data $b(t)=b$ and $p(t)=p$, then the
integrand in $F(b,p,t)$ will depend explicitly on time only through $b$ and
$p$. This means that the time dependence of $U(t^{\prime })$ may be written
as $U(b,p,t^{\prime }-t)$. If we now write the Cauchy data in terms of the
action-angle variables ($J,\theta $), since the equation of motion for the
angle variable is simply $\dot \theta =\Omega $ we may write $b[B(\theta
,J),P(\theta ,J),t]=b(\theta +\Omega t,J)$ and similarly for $p$. This means
that we may substitute the time derivative operator $d/dt$ by $\Omega
\partial /\partial \theta $ in $F(b,p,t)$.

Thus substituting (\ref{51}) and (\ref{52}) into (\ref{b4}), using $\Omega
d/d\theta$ instead of $d/dt$, integrating by parts and using the expression
for $U(t)$ given by (\ref{48}) we get
\begin{equation}
S=\frac{6\nu}{2\pi\Omega}\int_0^{2\pi} d\theta \left[\frac{d}{dt}
\left(\frac{\dot  p}{b}\right)\right] I(t).  \label{b8}
\end{equation}

This may be simplified using the equation of motion (\ref{50}) to lowest
order, then changing $d\theta$ by $\Omega dt$ we have
\begin{equation}
S=-\frac{\nu^2\Lambda^2}{2\pi}\int_0^{2\pi/\Omega} dt \left(
\frac{d}{dt} b^2(t)\right)\int_{-\infty}^\infty dt^\prime
H(t-t^\prime)b^2(t^\prime). \label{b9}
\end{equation}
Note that this term is of order $\nu^2\Lambda^2$ as the
diffusion term (\ref{b1}). Thus it is convenient to introduce ${\bf S}$ by
\begin{equation}
S(J)=\frac{\pi \nu ^2\Lambda ^2}4{\bf S}(J).  \label{b10}
\end{equation}
Now we can make use of (\ref{53d}) for the kernel $H$ (note that the local
delta term does not contribute), and introduce a new variable 
$u=t-t^\prime$, instead of $t^\prime$ to write ${\bf S}$ as
\begin{equation}
{\bf S}(J)=\frac{-1}{4\pi ^2}\int_0^{2\pi /\Omega }dt\left( 
\frac d{dt}b^2(t)\right) {\rm Pf}\int_0^\infty \frac{du}ub^2(t-u).
\label{b11}
\end{equation}

The equation for the stationary flux (\ref{b7}) becomes
\begin{equation}
\frac{{\bf D}(J)}\Omega \frac{\partial f}{\partial J}-{\bf S}(J)f=
-\frac 4{\pi \nu ^2\Lambda ^2}K_0.  \label{b12}
\end{equation}
All that remains now is to find appropriate expressions for ${\bf D}$ and
${\bf S}$ in this equation and follow Kramers' problem in the Appendix to
compute $K_0$. From now on, however, it is more convenient to use the energy
$E$ as a variable instead of $J$, where $E=H(J)$ and thus we will compute 
${\bf D}(E)$ and ${\bf S}(E)$ in what follows.

\subsection{Evaluating ${\bf S}$ and ${\bf D}$}

Let us begin by recalling the basic features of the classical orbits. The
most important feature of the classical dynamics is the presence of two
unstable fixed points at $p=0$, $b=\pm 2\sqrt{E_s}$, where $E_s=3/(2\Lambda) $
is also the corresponding value of the ``energy" $E=p^2/2+V\left( b\right)$.
These fixed points are joined by a heteroclinic orbit or separatrix.
Motion for energies greater than $E_s$ is unbounded. For $E\leq E_s$, we
have outer unbound orbits and inner orbits confined within the potential
well. These periodical orbits shall be our present concern.

As it happens, the orbits describing periodic motion may be described in
terms of elliptic functions (see Appendix D). The exact expression for the
orbits leads to corresponding expressions for ${\bf D}$ and ${\bf S}$ (see
Appendix E). Introducing a variable $k$
\begin{equation}
k^2=\frac{1-\sqrt{1-\frac E{E_s}}}{1+\sqrt{1-\frac E{E_s}}},  \label{new1}
\end{equation}
so that $k^2\sim E/4E_s$ for low energy, while $k^2\rightarrow 1$ as we
approach the separatrix, we find
\begin{equation}
{\bf D}\left( E\right) =\left( \frac{8E^2}{15\pi }\right) \frac{\left(
1+k^2\right) ^{3/2}}{k^4}\left\{ 2\left( 1-k^2+k^4\right) E\left[ k\right]
-\left( 2-3k^2+k^4\right) K\left[ k\right] \right\},  \label{new2}
\end{equation}
where $K$ and $E$ are the complete elliptic integrals of the first and
second kind (see \cite{abra}, \cite{WW})
\begin{equation}
K\left[ k\right] =\int_0^1\frac{dx}{\sqrt{\left( 1-x^2\right) \left(
1-k^2x^2\right) }},\ \ \ \ \;E\left[ k\right] =
\int_0^1dx\;
\sqrt{\frac{\left(1-k^2x^2\right) }{\left( 1-x^2\right) }}. 
\label{new3} 
\end{equation}

The corresponding expression for ${\bf S}$ is
\begin{equation}
{\bf S}\left( E\right) =\left( \frac{8E^2}{\pi ^2}\right) \frac{\left(
1+k^2\right) ^2}{k^4}\left\{ \alpha \left[ k\right] E\left[ k\right] -\gamma
\left[ k\right] K\left[ k\right] \right\},  \label{new5}
\end{equation}
where
\begin{equation}
\alpha \left[ k\right] =\int_0^\infty \frac{du}{u^2{\rm sn}^2u}\left\{
1-\left( \frac{1+k^2}3\right) {\rm sn}^2u-\left( \frac u{{\rm sn\,}u}\right)
\left[ 1-\left( 1+k^2\right) {\rm sn}^2u+
k^2{\rm sn}^4u\right] ^{1/2}\right\},
\label{new6}
\end{equation}
${\rm sn}\,u$ being the Jacobi elliptic function, and
\begin{equation}
\gamma \left[ k\right] =\int_0^\infty \frac{du}{u^2{\rm sn}^2u}\left\{
1-\left( \frac{1+2k^2}3\right) {\rm sn}^2u-\left( 
\frac{E\left[ u,k\right] }{{\rm sn\,}u}\right) \left[ 
1-\left( 1+k^2\right) {\rm sn}^2u+k^2{\rm sn}^4u\right] ^{1/2}\right\}, 
\label{new7}
\end{equation}
where $E\left[ u,k\right] $ is the incomplete elliptic integral of the
second kind
\begin{equation}
E\left[ u,k\right] =\int_0^{{\rm sn}\,u}dx\;\sqrt{\frac{\left(
1-k^2x^2\right) }{\left( 1-x^2\right) }}.  \label{new8}
\end{equation}

The conclusion of all this is that, while ${\bf D}$ and ${\bf S}$
individually behave as $E^2$ times a smooth function of $E/E_s$, their ratio
is relatively slowly varying. At low energy, we find ${\bf D\sim }E^2/2$ and 
${\bf S\sim }E^2/4$. As we approach the separatrix, 
${\bf D\rightarrow } 0.96\,E_s^2$ and ${\bf S\rightarrow }1.18\,E_s^2$.
Meanwhile, the ratio of the two goes from $0.5$ to $1.23$.

This means that we can write the equation for stationary distributions as
\begin{equation}
\frac{\partial f}{\partial E}-\beta \left( E\right) f=-\frac 4{\pi \nu
^2\Lambda ^2g\left( E\right) }\left( \frac{K_0}{E^2}\right),  \label{b12b}
\end{equation}
where $\beta $ and $g$ are smooth order one functions. There is a
fundamental difference with respect to Kramers' problem, namely the sign of
the second term in the left hand side. In the cosmological problem, the
effect of nonlocality is to favour diffussion rather than hindering it. We
may understand this as arising from a feedback effect associated with
particle creation (see \cite{Parker}).

\section{The tunneling amplitude}
\label{section4}

Having found the reduced Fokker-Planck equation Eq. (\ref{b12b}), we must
analyze its solutions in order to identify the range of the flux $K_0$. We
shall first consider the behavior of the solutions for $E\leq E_s$, and then
discuss the distribution function beyond the separatrix. Since our
derivation is not valid there, for this later part we will have to return to
an analysis from the equations of motion.
For concreteness, in what follows it is convenient to choose the order of 
magnitude of the cosmological constant. We shall assume a model geared to
produce GUT scale inflation, thus $\Lambda\sim 10^{-12}$, and
correspondingly $E_s\sim 10^{12}$ is very large in natural units.

\subsection{Distribution function inside the potential well}

As we have already discussed, the approximations used in building
our model break down at the cosmological singularity, and therefore Eq.
(\ref{b12b}) cannot be assumed to hold in a neighborhood of $E=0$. Thus it is
best to express the solution for $f$ in terms of its value at $E=E_s$
\begin{equation}
f(E)=\frac {4K_0}{\pi \nu^2\Lambda ^2 }\left[\sigma e^{\int^E dE'\beta(E')}
+f_p(E)\right],
\end{equation}
where $\sigma$ is an arbitrary constant and the particular solution 
$f_p(E)$ is chosen to vanish at $E=E_s$
\begin{equation}
f_p(E)= e^{\int^E dE'\beta(E')}\int_E^{E_s}
\frac{dE'}{g(E')E'^2}e^{-\int^{E'} dE''\beta(E'')}, 
\end{equation}
so that
\begin{equation}
f(E_s)\sim \frac {4K_0}{\pi \nu^2\Lambda ^2 }\sigma e^{\beta(E_s) E_s}.
\end{equation}

Because of the exponential suppression, the particular solution  
is dominated by the lower limit in the integral, leading to
\begin{equation}
f_p(E)\sim\frac{1}{g(E)E^2[\beta(E)+2/E]}-
\frac{e^{-\beta(E_s)(E_s-E)}}{g(E_s)E_s^2[\beta(E_s)+2/E_s]}.
\end{equation}

For $E\ll 1$ we see that $f_p\sim E^{-1}$, but
this behavior cannot be extrapolated all the way to zero as it
would make $f$ non integrable. However we must notice that neither our
treatment (i.e., the neglect of logarithmic potential corrections) nor
semiclassical theory generally is supposed to be valid arbitrarily close to
the singularity. Thus we shall assume that the pathological behavior of Eq.
(\ref{b12b}) near the origin will be absent in a more complete theory, and
apply it only from some lowest energy $E_\delta\sim 1 $ on. There are still
12 orders of magnitude between $E_\delta$ and $E_s$.

Since we lack a theory to fix the value of the constant $\sigma$, we shall
require it to be generic in the following sense. We already know that $f_p$
vanishes at $E_s$, by design, and then from the transport equation
(\ref{b12b}) we derive
$df_p/dE=-[g(E_s)E_s^2]^{-1}$ there. So unless 
$\sigma\leq [\beta(E_s)g(E_s)E_s^2]^{-1}\exp[-\beta(E_s) E_s]\sim
10^{-24}\exp(-10^{12})$,  $f$ has a positive slope as it approaches the
separatrix from below. We shall assume a generic $\sigma$ as one much above this
borderline value, so that for $E\geq 1$ the right hand side of the reduced
Fokker-Planck equation may be neglected, and $f$ grows exponentially
\begin{equation}
f(E) \sim \frac{4K_0\sigma}{\pi\nu^2\Lambda^2}e^{\beta(E)E}
\label{highen}
\end{equation}

\subsection{Outside the well}

Beyond the separatrix, all motion is unbounded and there is no analog of
action-angle variables, so we must return to the original variables $b$,
$p$. Also note that we are only interested in the regime when $E\geq E_s$,
that is, we shall not consider unbound motion below the top of the potential.

Let us first consider the behavior of classical orbits in the $(b,p)$ plane.
Our first observation is that as the universe gets unboundedly large, the
effects of spatial curvature become irrelevant. This means that we may
approximate $U\sim -6\nu \ddot b/b$, and accordingly the classical equation
of motion as $\ddot b\sim \Lambda b^3/6$.

In this regime, classical orbits are quickly drawn to a de Sitter type
expansion, whereby they can be parametrized as
\begin{equation}
b\left( t^{\prime }\right) =\frac{b\left( t\right) }{1+
\sqrt{\frac \Lambda {12}}b\left( t\right) (t-t^{\prime })}. 
\label{desitter} 
\end{equation}

After substituting $U\propto b^2$, it is easily seen that the nonlocal term
$I$ is proportional to $b^2(t)$, and that therefore the nonlocal force $F$
vanishes (see Eq. (\ref{51})). Therefore what we are dealing with are the local
quantum fluctuations of the metric, which one would not expect to act in a
definite direction, but rather to provide a sort of diffussive effect. To see
this, let us observe that if we look at the Fokker-Planck equation as a
continuity equation, then we may write it as
\[
\frac{\partial f}{\partial t}=-\vec \nabla \vec K, 
\]
and this allows us to identify the flux. For example, if the Fokker-Planck
equation reads
\[
\frac{\partial f}{\partial t}=\frac{\partial A}{\partial b}+
\frac{\partial B}{\partial p}, 
\]
then whatever $A$ and $B$ are, $\vec K=-A\hat b-B\hat p$, where a circunflex
denotes an unit vector in the corresponding direction. Rather than $\hat b$
and $\hat p$, however, it is convenient to use the components of $\hat K$
along and orthogonal to a classical trajectory. Since the energy $E$ is
constant along trajectories, $\vec \nabla E$ lies in the orthogonal
direction, so the orthogonal component is simply $K_E$, or, since $E=H(J)$,
$K_J$.

Our whole calculation so far amounts to computing the mean value of $K_J$
(see eq. (\ref{b6})); indeed the first term acts as diffussion, opposing the
gradients of $f$. The big surprise is the second term being positive,
forcing a positive flux towards larger energies.
Observe that, in particular, the mean flux across the separatrix is
positive. Since for a stationary solution the flux is conserved, the flux
must be positive accross any trajectory. Now beyond the separatrix the
term $S$ of (\ref{b6}) is absent because $F$ vanishes and, as we shall see,
${\bf D}$ remains positive. So, to obtain a positive flux, it is necessary that
$\partial f/\partial E<0,$ as we will now show.

To compute ${\bf D}$ beyond the separatrix, we observe that although there
are no longer action-angle variables, we may still introduce a new pair of
canonical variables $\left( E,\tau \right) $, where $E$ labels the different
trajectories and $\tau $ increases along classical trayectories, with $\dot 
\tau =1$. It works as follows. The relationship between $p$ and $b$,
$p=\sqrt{2E+\frac \Lambda {12}b^4}$, becomes, for low energy
\begin{equation}
p=\sqrt{\frac \Lambda {12}}b^2+\sqrt{\frac{12}\Lambda }\frac E{b^2}.
\end{equation}
This same relationship corresponds to a canonical transformation with
generating functional $W$,
\[
W\left( b,E\right) =\sqrt{\frac \Lambda {12}}\frac{b^3}3-
\sqrt{\frac{12}\Lambda }\frac Eb,
\]
and the new canonical coordinate $\tau $ follows from
\begin{equation}
\tau =\frac{\partial W}{\partial E}=-\sqrt{\frac{12}\Lambda }\frac 1b.
\end{equation}
Comparing with Eq. (\ref{desitter}), this is just
\[
\tau =-\sqrt{\frac{12}\Lambda }\frac 1{b( t_0) }\left[ 1+
\sqrt{\frac \Lambda {12}}b( t_0)( t_0-t) \right], 
\]
for some constant of integration $t_0$. Indeed $\dot \tau =1$, as it must.

Writing the Fokker-Planck equation (\ref{FokkerPlanck}) in the new variables
$\left( E,\tau \right) $ is an exercise in Poisson brackets, simplified by the
approximation $\partial b/\partial E\sim 0$ (to see that this approximation is
justified we may go to one more order in $E$ in the expressions for $p$, $W$ and
$\tau$ and we find that for large $b$, $\partial b/\partial E\sim -12/(5\Lambda
b^3)$). Thus from
Eq. (\ref{FokkerPlanck}) with $F=0$, we get
\begin{equation}
\frac{\partial f}{\partial t}=-\frac{\partial f}{\partial \tau }+\frac{\pi
\nu ^2\Lambda ^3}{48}b^6\frac{\partial ^2f}{\partial E^2},
\label{nfp}
\end{equation} 
so that $K_\tau =f$ (that is, the universe moves along the classical trajectory
with $\dot \tau =1$), and
\[
K_E=-\frac{\pi \nu ^2\Lambda ^3}{48}b^6\frac{\partial f}{\partial E}. 
\]
with only the normal diffussive term present, as it was expected. Since $K_E$
must be positive (at least in the average) $f$ must decreases beyond the
separatrix, as we wanted to show.

This result, in fact, can be made more quantitative if we note that Eq.
(\ref{nfp}) for a stationary distribution function $f$ is essentially a heat
equation which can be solved in the usual way. For this it is convenient to
change to a new variable $s=-1/(5\tau^5)$ which is positive semidefinite since
the conformal time $\tau$ is negative in the de Sitter region. The equation then
can be written as,
\begin{equation}
\frac{\partial f}{\partial s}=d\frac{\partial^2 f}{\partial E^2},
\label{heat}
\end{equation}
where $d\equiv 36\pi\nu^2$. Its solution can be written as
\begin{equation}
f(E)=\frac{1}{\sqrt{4\pi d s}}\int dE' e^{\frac{(E-E')^2}{4sd}}h(E'),
\label{heat1}
\end{equation}
where $h(E')$ is a function which determines the value of $f$ at
$\tau=-\infty$. It is easy to compute
\begin{equation}
\int_{-\infty}^0 d\tau f(E,\tau)\propto
\int_0^\infty dE'\frac{h(E')}{(E-E')^{7/5}},
\label{heat2}
\end{equation}
which shows that for large $E$, $f$ in fact decreases as $E^{-7/5}$.

\subsection{The tunneling amplitude}

After the two previous subsections, we gather that the stationary solutions
to the Fokker-Planck equation display a marked peak at $E=E_s$. We may now
estimate the flux by requesting, as we do for Kramers' problem in the
Appendix, that the total area below the distribution function should not
exceed unity. Unless the lower cutoff $E_\delta $ is very small (it ought to
be exponentially small on $E_s$ to invalidate our argument) the integral is
dominated by that peak, and we obtain
\begin{equation}
K_0\leq ( {\rm prefactor}) \exp\left[ -\beta ( E_s)
E_s\right]. 
\label{tuna}
\end{equation}
The prefactor depends on $\Lambda $, $\nu $, $g\left( 1\right) $, $\beta
\left( 1\right) $, $\sigma$ and the details of the peak shape. Using 
$E_s=3/(2\Lambda) $, $\beta \left( E_s\right) =1.23$, we get
\begin{equation}
K_0\leq ( {\rm prefactor}) \exp\left( -\frac{1.84}\Lambda \right).
\label{tunb}
\end{equation}
In the last section of the Appendix we have computed the flux when one
considers a cosmological model with a single cosmic cycle. The result
(\ref{finalflux}) is qualitatively similar to this one, it just gives a sligthly
lower probability. This semiclassical result must now be compared against the
instanton calculations.

\section{Conclusions}
\label{section5}

In this paper we have studied the possibility that a closed isotropic
universe trapped in the potential well produced by a cosmological constant
may go over the potential barrier as a consequence of back-reaction to the
quantum effects of a non conformally coupled quantum scalar field. The
quantum fluctuations of this field act on geometry through the stress-energy
tensor, which has a deterministic part, associated to vacuum polarization
and particle creation, and also a fluctuating part, related to the
fluctuation of the stress-energy itself. The result is that the scale
factor of the classical universe is subject to forcing due to particle
creation and also to a stochastic force due to these fluctuations. We
compute the Fokker-Planck equation for the probability distribution of the
cosmological scale factor and compute the probability that the scale factor
crosses the barrier and ends up in the de Sitter stage where 
$b\sim\sqrt{12/\Lambda }\cosh (\sqrt{\Lambda /12}t^{\prime })$, where 
$t^{\prime}$ is cosmological time $bdt=dt^{\prime }$, if it was initially near
$b\sim 0$. The result displayed in (\ref{tunb}) is that such probability is
\begin{equation}
K_0\sim \exp \left( -\frac{1.8}\Lambda \right) ,  \label{b24}
\end{equation}
or a similar result, displayed in (\ref{finalflux}), if we consider a
cosmological model undergoing a
single cosmic cycle. This result is comparable with the probability that the
universe tunnels quantum mechanically into the de Sitter phase from nothing
\cite{Vilenkin 1982-1984}. In this case from the classical action (\ref{47}) 
$S_{g,m}^R[b]$, i.e. neglecting the terms of order $\nu $, one constructs the
Euclidean action $S_E$, after changing the time $t=i\tau $,
\begin{equation}
S_E[b]=\frac{1}{2}\int d\tau \left[\dot b^2+b^2-\frac{1}{12}\Lambda
b^4\right].  \label{b25}
\end{equation}
The Euclidean trajectory is $b=\sqrt{12/\Lambda}\cos(\sqrt{\Lambda/12}
\tau^{\prime})$, where $\tau^{\prime}$ is Euclidean cosmological time (this
is the instanton solution). This trajectory gives an Euclidean action
$S_E=4/\Lambda$. The tunneling probability is then
\begin{equation}
p\sim\exp \left(-\frac{8}{\Lambda}\right).  \label{b26}
\end{equation}

This result, which in itself is a semiclassical result, is comparable to
ours, (\ref{b24}), but it is of a very different nature. We have ignored the
quantum effects of the cosmological scale factor but we have included the
back-reaction of the quantum fields on this scale factor. Also our universe
reaches the de Sitter stage with some energy due to the particles that have
been created. In the instanton solution only the tunneling amplitude of the
scale factor is considered and the universe reaches the de Sitter phase with
zero energy.

Taken at face value, our results seem to imply that the nonlocality and
randomness induced by particle creation are actually as important as the
purely quantum effects. This conclusion may be premature since after all Eq.
(\ref{b24}) is only an upper bound on the flux. Nevertheless, our results
show that ignoring back-reaction of matter fields in quantum cosmology may
not be entirely justified. We expect to delve further on this subject in
future contributions.

\begin{center}
{\normalsize {\bf ACKNOWLEDGMENTS}}
\end{center}

We are grateful to Leticia Cugliandolo, Miquel Dorca, Larry Ford, Jaume Garriga,
Bei-Lok Hu, Jorge Kurchan, Diego Mazzitelli, Pasquale Nardone, Juan Pablo Paz,
Josep Porr\`a, Albert Roura and Alex Vilenkin for 
very helpful suggestions and discussions. This work has been partially
supported by the European project CI1-CT94-0004 and by the CICYT contracts
AEN95-0590, Universidad de Buenos Aires, CONICET and Fundaci\'on Antorchas.

\section{Appendix}
\label{section6}

To facilitate the reading of this Appendix we repeat here the summary of its
contents given in the Introduction: Appendix A gives
some details of the renormalization of the CTP effective action; Appendix B
explains how to handle the diffusion terms when the Fokker-Planck equation is
constructed; in Appendix C we formulate and discuss Kramers problem in
action-angle variables; the short Appendix D gives the exact classical 
solutions for the cosmological scale factor; in Appendix E the averaged
diffusion and dissipation coefficients for the averaged Fokker-Planck equation
are derived; in Appendix F the relaxation time is computed in detail; and
finally in Appendix G the calculation of the escape probability for the scale
factor is made for a model which undergoes a single cosmic cycle.

\subsection{Divergences of $T$}

Here we compute the finite imaginary part of the series defined in 
(\ref{32}) and prove that the real part diverges like $1/(n-4)$. 
The finite real part of the series will not be found explicitly, its exact
form is not needed in the calculation of this paper. Let us now call
$\varepsilon\equiv n-4$, and call $F(\omega)$ the series (\ref{32}) which we
can write in terms of the Gamma functions as,
\begin{eqnarray}
F(\omega ) &\equiv &\sum_{k=1}^\infty a_k(\omega )=\sum_{k=1}^\infty 
{\frac{\Gamma (k+\varepsilon +1)}{\Gamma (k)}}{\frac 1{(k+\varepsilon
/2)^2-(\omega /2)^2+i0^{+}}}  \nonumber \\
&=&\sum_{k=1}^\infty {\frac{\Gamma (k+\varepsilon +1)}{\Gamma (k)}}\left\{ 
{\rm PV}{\frac 1{(k+\varepsilon /2)^2-(\omega /2)^2}}-i\pi \delta
[(k+\varepsilon /2)^2-(\omega /2)^2]\right\}  \nonumber \\
&\equiv &F_R+iF_I,  \label{33}
\end{eqnarray}
where we have used that 
$(x\pm i0^{+})^{-1}={\rm PV}(1/x)\mp i\pi \delta (x)$. Let us first concentrate
in the imaginary part $F_I$ and compute, according to (\ref{31}), its Fourier
transform
\begin{equation}
\tilde F_I\equiv \int_{-\infty }^\infty {\frac{d\omega }{2\pi }}e^{-i\omega
(t-t^{\prime })}F_I={\frac 12}\sum_{k=1}^\infty {\frac{\Gamma (k+\varepsilon
+1)}{\Gamma (k)}}{\frac{\cos (k+\varepsilon /2)(t-t^{\prime })}
{k+\varepsilon /2}},  \label{34}
\end{equation}
where we have used that $2 (k+\varepsilon/2)\delta[(k+\varepsilon/2)^2-
(\omega/2)^2] =\delta(k+\varepsilon/2+\omega/2)+
\delta(k+\varepsilon/2-\omega/2)$. Now the last expression is clearly
convergent when $\varepsilon=0$, thus we get
\begin{equation}
\tilde F_I={\frac 12}\sum_{k=1}^\infty \cos k(t-t^{\prime }),  \label{35}
\end{equation}
this series can be summed up and we get the train of deltas of (\ref{29b}),
thus recovering the noise kernel, from $\tilde F_I=8N(t-t^{\prime })$.

Let us now see that the real part of the series diverges like 
$1/\varepsilon$. Using that $\Gamma(x+1)=x\Gamma(x)$ the principal part of
$a_k$ can also be written as
\begin{equation}
a_k(\omega)={\frac{\Gamma(k+\varepsilon)}{\Gamma(k)}} {\frac{k+\varepsilon 
} {(k+\varepsilon/2)^2-(\omega/2)^2}}.  \label{36}
\end{equation}

It is clear from this expression that the divergences when $\varepsilon=0$
come from the ratio of gamma functions in (\ref{36}) when $k$ is large. Let
us now separate the sum $\sum_{k=1}^\infty a_k=\sum_{k=1}^{N-1} a_k
+\sum_{k=N}^\infty a_k$ where $N\gg 1$. We can use now that for large $x$,
$\Gamma(x)=\sqrt{2\pi}x^{x-1/2} e^{-x}(1+O(1/x))$ and the definition of 
$e$, $e=\lim_{n\rightarrow\infty} (1+1/n)^n$, to prove that 
$\Gamma(k+\varepsilon)/\Gamma(k)= k^\varepsilon (1+O(1/k))$. 
Substituting $a_k$ by $\bar a_k$, defined by
\begin{equation}
\bar a_k=k^\varepsilon \left[ 1+O\left( {\frac{1}{k}}\right)\right] 
{\frac{k+\varepsilon}{(k+\varepsilon/2)^2-(\omega/2)^2}},  \label{37}
\end{equation}
in the second sum of the previous separation and we can write
$\sum_{k=1}^\infty a_k= \sum_{k=1}^{N-1} a_k+\sum_{k=N}^\infty \bar a_k$. Now
we can use the Euler-Maclaurin summation formula \cite{Gradshteyn 1979} to
write $\sum_{k=N}^\infty \bar a_k=\int_N^\infty dk \bar a_k+\dots$, where
the dots stand for terms which are finite since they depend on succesive
derivatives of $\bar a_k$ at the integration limits. Thus we may write 
$\sum_{k=1}^\infty a_k = \sum_{k=1}^{N-1} a_k -\int_0^N dk\, \bar a_k
+\int_0^\infty dk\, \bar a_k $. The first sum and first integral of this
last equation are finite for all $\varepsilon$, thus we can take 
$\varepsilon=0$, in which case $a_k=\bar a_k= k/(k^2-(\omega/2)^2)$. The sum
and integral may then be performed (writing $2a_k=
1/(k+\omega/2)+1/(k-\omega/2)$) and the $\ln N$ which appears in both
expressions cancel, the next to leading order terms differ by order
$O(1/N)$. Therefore the divergence is in the last integral
\begin{equation}
\int_0^\infty dk \bar a_k= \int_0^\infty dk {\frac{k^{\varepsilon +1}}
{(k+\varepsilon/2)^2-(\omega/2)^2}},  \label{38}
\end{equation}
where here $\varepsilon$ is an arbitrary parameter. This integral is easily
computed \cite{Gradshteyn 1979}, and when it is expanded in powers of
$\varepsilon$ we get
\begin{equation}
\int_0^\infty dk \bar a_k=-\left[ {\frac{1}{n-4}}+{\frac{1}{2}}\ln
(\omega/2)^2 \right].  \label{39}
\end{equation}

Thus, according to (\ref{31}), (\ref{32}) and (\ref{33}) we compute the
Fourier transform of $F_R$,
\begin{equation}
\tilde F_R= -{\frac{\delta (t-t^{\prime})}{n-4}}-8 A(t-t^{\prime}),
\label{39bis}
\end{equation}
where $A(t-t^{\prime})$ stands for a finite kernel (see Eq. (\ref{53c}) ). 

\subsection{The diffussion terms}

We want to compute (\ref{54b}) which can be written as (\ref{Phi}) in terms of
the functions $C_n$ ($n=0,1,2$) of (\ref{C_i}).
The simplest function $C_0$ can be written after using (\ref{novikov}) as,
\[
C_0=\int dt^{\prime }\;N( t-t^{\prime }) \left\langle\frac \delta {\delta
\xi \left( t^{\prime }\right) }\delta \left( b\left( t\right) -b\right)
\delta \left( p\left( t\right) -p\right) \right\rangle,
\]
whereas to write the other two functions we observe that the noise kernel is
translation invariant, so integrating by parts (in a distribution sense)
\[
C_n=\int dt^{\prime }\;N\left( t-t^{\prime }\right) \frac{\partial ^n}
{\partial t^{\prime n}}\left\langle
\frac \delta {\delta \xi 
\left( t^{\prime }\right) }\delta \left( b\left( t\right) -b\right) \delta
\left( p\left( t\right) -p\right) \right\rangle. 
\]

We now use the local approximation for the noise kernel to get
\[
C_0=\frac \pi {16}\left. \left\langle 
\frac \delta {\delta \xi \left( t^{\prime }\right) 
}\delta \left( b\left( t\right) -b\right) \delta \left( p\left( t\right)
-p\right) \right\rangle\right| _{t^{\prime }\rightarrow t} ,
\]
and similarly $C_1$ and $C_2$. As we know, this reduces to
\[
C_0=\frac{-\pi }{16}\left( \frac \partial {\partial b}
\left\langle\frac{\delta b\left(
t\right) }{\delta \xi \left( t^{\prime }\right) }\delta \left( b\left(
t\right) -b\right) \delta \left( p\left( t\right) -p\right) 
\right\rangle+\frac \partial
{\partial p}\left\langle
\frac{\delta p\left( t\right) }{\delta \xi \left( t^{\prime
}\right) }\delta \left( b\left( t\right) -b\right) \delta \left( p\left(
t\right) -p\right) \right\rangle \right) .
\]

A functional derivative of the equations of motion leads to
\begin{eqnarray*}
\frac d{dt}\frac{\delta b\left( t\right) }{\delta \xi \left( t^{\prime
}\right) } &=&\frac{\delta p\left( t\right) }{\delta \xi \left( t^{\prime
}\right) }, \\
\frac d{dt}\frac{\delta p\left( t\right) }{\delta \xi \left( t^{\prime
}\right) } &=&-V^{\prime \prime }\left[ b\left( t\right) \right] 
\frac{\delta b\left( t\right) }{\delta \xi \left( t^{\prime }\right) }+6\nu
\left[  \frac 1{b\left( t^{\prime }\right) }\frac{d^2}{dt^2}\delta \left(
t-t^{\prime }\right) +\frac{V^{\prime }\left[ b\left( t^{\prime }\right)
\right] }{b^2\left( t^{\prime }\right) }\delta \left( t-t^{\prime }\right)
\right],
\end{eqnarray*}
where actually we are computing the right hand side only to lowest order in
$\nu$. This suggests writing
\begin{eqnarray}
\frac{\delta p\left( t\right) }{\delta \xi \left( t^{\prime }\right) }
&=&G\left( t-t^{\prime }\right) \theta \left( t-t^{\prime }\right) +
\frac{6\nu }{b\left( t^{\prime }\right) }\frac d{dt}\delta \left( t-t^{\prime
}\right),  \label{nine} \\
\frac{\delta b\left( t\right) }{\delta \xi \left( t^{\prime }\right) }
&=&R\left( t-t^{\prime }\right) \theta \left( t-t^{\prime }\right) +
\frac{6\nu }{b\left( t^{\prime }\right) }\delta \left( t-t^{\prime }\right), 
\nonumber
\end{eqnarray}
which works provided
\begin{eqnarray}
\frac{dR}{dt} &=&G,\ \ \ \ R\left( 0\right) =0,  \nonumber \\
\frac{dG}{dt} &=&-V^{\prime \prime }\left[ b\left( t\right) \right]
R,\ \ \ \ G\left( 0\right) =6\nu \left[ \frac{V^{\prime }
\left[ b\left( t^{\prime}\right) \right] }{b^2\left(
t^{\prime}\right)}-\frac{V^{\prime \prime }\left[ b\left( t^{\prime }\right)
\right] } {b\left( t^{\prime }\right) }\right] =2\nu \Lambda b(t^{\prime }).  
\nonumber \end{eqnarray}

In the coincidence limit
\begin{equation}
\left. \frac{\delta b\left( t\right) }{\delta \xi 
\left( t^{\prime }\right) }\right| _{t^{\prime }\rightarrow t}=0,\ \ \ \ 
\left.\frac{\delta p\left( t\right)  }{\delta \xi \left( t^{\prime }\right)
}\right| _{t^{\prime }\rightarrow t}=2\nu \Lambda b ,  
\label{eleven}
\end{equation}
which leads to
\begin{equation}
\left.\left\langle 
\frac \delta {\delta \xi \left( t^{\prime }\right) }\delta \left(
b\left( t\right) -b\right) \delta \left( p\left( t\right) -p\right) 
\right\rangle\right|
_{t^{\prime }\rightarrow t}=-2\nu \Lambda b
\frac \partial {\partial p}f\left( b,p,t\right).  
\label{twelve}
\end{equation}

The diffusive terms also involve the first and second derivatives of the
propagators with respect to $t^{\prime }$. To find them, we make the following
reasoning. We have just seen that, for example, $R(t,t)\equiv 0$, therefore
\begin{equation}
\left. \frac \partial {\partial t^{\prime }}R\left( t,t^{\prime }\right)
\right| _{t^{\prime }\rightarrow t}=-\left. 
\frac \partial {\partial t}R\left( t,t^{\prime }\right) \right| _{t^{\prime
}\rightarrow t}=-G\left( t,t\right) =-2\nu \Lambda b.  \label{fourteen}
\end{equation}

With a slight adaptation, we also get
\[
\left. \frac \partial {\partial t^{\prime }}G\left( t,t^{\prime }\right)
\right| _{t^{\prime }\rightarrow t}=\frac \partial {\partial t}\left[ \left.
G\left( t,t^{\prime }\right) \right| _{t^{\prime }\rightarrow t}\right]
-\left. \frac \partial {\partial t}G\left( t,t^{\prime }\right) \right|
_{t^{\prime }\rightarrow t} ,
\]
so that we have
\begin{equation}
\left. \frac \partial {\partial t^{\prime }}G\left( t,t^{\prime }\right)
\right| _{t^{\prime }\rightarrow t}=2\nu \Lambda p .
\label{fifteen}
\end{equation}
Iterating this argument, we find
\[
\left. \frac{\partial ^2}{\partial t^{\prime 2}}R\left( t,t^{\prime }\right)
\right| _{t^{\prime }\rightarrow t}=-\frac \partial {\partial t}G\left(
t,t\right) -\left. \frac \partial {\partial t^{\prime }}G\left( t,t^{\prime
}\right) \right| _{t^{\prime }\rightarrow t} ,
\]
where we have permutted a $t$ and a $t^{\prime }$ derivative and used the
equations of motion. From this we thus get
\begin{equation}
\left. \frac{\partial ^2}{\partial t^{\prime 2}}R\left( t,t^{\prime }\right)
\right| _{t^{\prime }\rightarrow t}=-4\nu \Lambda p  .
\label{sixteen}
\end{equation}

The last formula of this type that we need is
\[
\left. \frac{\partial ^2}{\partial t^{\prime 2}}G\left( t,t^{\prime }\right)
\right| _{t^{\prime }\rightarrow t}=\frac \partial {\partial t}\left[ \left. 
\frac \partial {\partial t^{\prime }}G\left( t,t^{\prime }\right) \right|
_{t^{\prime }\rightarrow t}\right] -\left. \frac \partial {\partial
t^{\prime }}\frac \partial {\partial t}G\left( t,t^{\prime }\right) \right|
_{t^{\prime }\rightarrow t} ,
\]
which from the equations of motion leads to
\begin{eqnarray}
\left. \frac{\partial ^2}{\partial t^{\prime 2}}G\left( t,t^{\prime }\right)
\right| _{t^{\prime }\rightarrow t} &=&-2\nu \Lambda V^{\prime }\left(
b\right) +\frac \partial {\partial t^{\prime }}V^{\prime \prime }\left[
b\left( t\right) \right] R\left( t,t^{\prime }\right)  \label{seventeen} \\
\ &=&-2\nu \Lambda V^{\prime }\left( b\right) -V^{\prime \prime }\left[
b\right] 2\nu \Lambda b  \nonumber \\
\ &=&-2\nu \Lambda \frac \partial {\partial b}\left( bV^{\prime }\left[
b\right] \right) .
\nonumber
\end{eqnarray}

\subsection{Kramers problem}

For our purposes in this paper we call Kramers problem \cite{Kramers} the
computation of the ``tunneling amplitude" or, more properly, the escape
probability of a particle confined in a potential $V(b)$, such as 
(\ref{two}) for instance, which has a maximum and a separatrix with an energy
$E_s$. The particle is subject to a damping force $\gamma p$ ($p=\dot b$) and
white noise with amplitude $\gamma k T$, according to the
fluctuation-dissipation relation, where $\gamma$ is a friction coefficient, $k$
Boltzmann constant and $T$ the temperature. The Fokker-Planck equation in this
case is  \cite{Risken 1984}
\begin{equation}
\frac {\partial f}{\partial t}= \{ H,f\} +\gamma\frac {\partial}{\partial p}
\left[ pf+kT \frac{\partial f}{\partial p}\right],  \label{a1}
\end{equation}
where $H$ is given by (\ref{54a}). Since the particle is trapped in the
potential it undergoes periodic motion, in this case it is convenient to
introduce action-angle variables \cite{Goldstein} ($J, \theta$) as canonical
variables instead of ($b,p$), thus making a canonical transformation
$b=B(\theta,J)$, $p=P(\theta,J)$. The action variable $J$ is defined by
\begin{equation}
J=\frac{1}{2\pi}\oint pdb.  \label{a2}
\end{equation}

Since $p$ can be written in terms of $b$ and $H$, substitution in (\ref{a2})
and inversion implies that $H=H(J)$, and
\begin{equation}
\frac{\partial H}{\partial J}=\Omega(J),  \label{a3}
\end{equation}
is the frequency of the motion. The other canonical variable, the angle
variable $\theta $, satisfies a very simple equation of motion $\dot \theta
=\Omega $ and changes from $0$ to $2\pi $. At high energies, that is, near
the separatrix when $J\rightarrow J_s$, the motion ceases to be periodic and 
$\Omega \rightarrow 0$. At low energies, let as assume that $b=0$ is a
stable minimum of the potential, near this minimum the potential approaches
the potential of a harmonic oscillator with frequency $\omega $, $V(b)\sim
\omega b^2/2$ (in our case we simply have $\omega =1$), then 
$J\rightarrow 0$, $H\sim \omega J$ and $\Omega \sim \omega $.

If $\gamma =0$, then the solution to the Fokker-Planck equation is an
arbitrary function of $J$ and $\theta -\Omega t$. Stationary solutions are
therefore functions of $J$ alone. We may seek a general solution as
\[
f\left( J,t\right) +\gamma \sum_{n\neq 0}c_n\left( J,t\right) 
e^{in\left(\theta -\Omega t\right) }, 
\]
in this case we have $\partial _pf=\partial _pJ|_b\partial _Jf$. From (\ref
{54a}) we have that $p=[2(H-V(b))]^{1/2}$ and, consequently, $\partial
_Jp|_b=\Omega /p$ whose inverse is $\partial _pJ|_b=p/\Omega $. This can be
used to write $\partial ^2f/\partial p^2$ in terms of derivatives with
respect to $J$, and since now $\{H,f(J)\}=0$ we can write the Fokker-Planck
equation (\ref{a1}) in the new variables as, keeping only first order terms,
\begin{equation}
\frac{\partial f}{\partial t}+\gamma \sum_{n\neq 0}\frac{\partial c_n}
{\partial t}\left( J,t\right) e^{in\left( \theta -\Omega t\right) }=\gamma
\left[ f+\frac{p^2}\Omega \frac{\partial f}{\partial J}+kT\left( 
\frac 1\Omega \frac{\partial f}{\partial J}+\frac{p^2}\Omega 
\frac \partial {\partial J}\left( \frac 1\Omega 
\frac{\partial f}{\partial J}\right) \right) \right].  \label{a4}
\end{equation}

Fourier expanding the coefficients in the right hand side we obtain a set of
equations for the $c_n$ coefficients. The equation for $f$ itself follows
from the average of this equation over the angle variable $\theta $. Let us
change the integration variable in the definition (\ref{a2}) of $J$,
$db=\partial _\theta b|_Jd\theta $, taking into account that over a classical
trajectory $J$ is constant, and that $\dot \theta =\Omega $ we have
$\partial _\theta b|_J=p/\Omega (J)$. Thus, we can write (\ref{a2}) as,
\begin{equation}
\frac{1}{2\pi}\int_0^{2\pi} d\theta p^2=J\Omega(J).  \label{a5}
\end{equation}

Using this result we can now take the average of equation (\ref{a4}) over
$\theta$. This average reads simply,
\begin{equation}
\frac{\partial f}{\partial t}=\gamma\frac{\partial}{\partial J} \left(
J\left[ f+\frac{kT}{\Omega}\frac{\partial f}{\partial J}\right]\right).
\label{a6}
\end{equation}

As one would expect $\exp (-E/kT)$ is a solution of this equation. Let us
now see whether this equation, which is a transport equation, admits
stationary solutions with positive probability flux \cite{Langer}. Note that we
may write this equation as a continuity equation $\partial _tf+\partial _JK=0$,
where the flux $K$ can be read directly from (\ref{a6}). Therefore a stationary
solution with positive flux $K_0$ should satisfy
\begin{equation}
\frac{kT}{\Omega}\frac{\partial f}{\partial J}+f=-\frac{K_0}{\gamma J},
\label{a7}
\end{equation}
which can be integrated to give,
\begin{equation}
f=\frac{K_0}{\gamma k T} e^{-E/kT}\int_J^{J_0}\frac{d\xi}{\xi}
\Omega(\xi)e^{E(\xi)/kT}.  \label{a8}
\end{equation}

For any $K_0$, $f$ diverges logarithmically when $J\rightarrow 0$, however
this is an integrable singularity in $J$ and this is not a problem as we
will see shortly. In our problem the action variable $J$ satisfies that
$J\leq J_s$ and equation (\ref{a8}) proves that there is a real and positive
solution for any $J$ in such a range, which corresponds to choosing 
$J_0=J_s$.

Given a solution we may determine the flux $K_0$ imposing the condition that
the probability of finding the particle trapped in the potential well should
not be greater than unity \cite{Langer}, i.e. $\int_0^{J_s}f(J)dJ\leq 1$. This is
equivalent to
\begin{equation}
1\geq\frac{K_0}{\gamma k T} \left(\int_0^{J_s}\frac{d\xi}{\xi}
\Omega(\xi)e^{E(\xi)/kT} \int_0^\xi dJ e^{-E/kT}\right)  \label{a9}
\end{equation}
Since the integral is regular at zero it is dominated by the contribution
from the upper limit, and the integral may be evaluated approximately. One
gets
\begin{equation}
K_0\leq \gamma \frac{\omega J_s}{kT} \exp\left(-\frac{E_s}{kT}\right),
\label{a10}
\end{equation}
where we have used that near the separatrix $H\sim \omega J_s$. Typically
the flux is very small so that the probability of finding the particle in
the potential well is nearly one, therefore the value of $K_0$ approaches
the right hand side of (\ref{a10}).
 
We should remark here that in the order
reduction scheme that we are following, to compute the noise and the
non local terms we use the
classical equations of motion. In fact, these terms have a
quantum origin in our case and its computation is one of the tasks we have to
perform in order to define our particular Kramers problem. Thus the use of the
action-angle variables, which is convenient for the classical equations of
motion, is also convenient (after order reduction) 
in our approach to the Kramers problem. 

\subsection{A look at the orbits}

In what follows, we shall quote extensively from Abramowitz and Stegun,
Ref. \cite{abra} (from now on, AS), and
Whittaker and Watson, Ref. \cite{WW} (henceforth, WW).

The motion is described by the Hamiltonian
\begin{equation}
H=\frac 12\left( p^2+b^2\right) -\frac \Lambda {24}b^4. 
\label{hamiltonian}
\end{equation}
The energy is conserved, and on an energy surface $H=E$, the momentum is
$
p^2=2E-b^2+\Lambda b^4/12 .
$
The classical turning points correspond to $p=0$. Introducing the separatrix
energy
$
E_s= 3/(2\Lambda ),  
$
we can write the four turning points as 
\begin{equation}
b_{\pm }^2=4E_s\left[ 1\pm \sqrt{1-\frac E{E_s}}\right],  \label{bplus}
\end{equation}
two of them $\pm b_{-}$ are inside the
barrier, and two $\pm b_{+}$ are outside it.
The momentum can now be written as
\begin{equation}
p^2=2E\left( 1-k^2\frac{b^2}{b_{-}^2}\right) \left( 1-\frac{b^2}{b_{-}^2}
\right),  \label{momentum2}
\end{equation}
where we have introduced $k^2=(b_-/b_+)^2$, see Eq. (\ref{new1}).
The equation for the orbit is $b=b_{-}x(t),$ where
\begin{equation}
x={\rm sn}\left[ \frac{b_{+}t}{\sqrt{8E_s}},k\right], \label{elliptic}
\end{equation}
where sn is the Jacobi Elliptic Function (we follow the notation from WW 22.11;
to convert to AS, put $m=k^2$, and see AS 16.1.5).

The Jacobi elliptic function is periodic with period $4K\left[ k\right] $,
where $K$ is the complete elliptic integral of the first kind (AS 16.1.1 and
17.3.1) (see Eq. (\ref{new3})). The period in physical time is 
$T=\sqrt{8E_s}4Kb_{+}^{-1}$, and the frequency
\begin{equation}
\Omega =\frac{\pi b_{+}}{2\sqrt{8E_s}K\left[ k\right] }.  \label{frequency}
\end{equation}

\subsection{The D and S functions}

The function ${\bf D}$ is given by
\[
{\bf D}(J)=\frac 1{2\pi }\int_0^{2\pi /\Omega }dt\;b^2p^2 ,
\]
which by introducing $b=b_{-}x$ can be written as
\begin{eqnarray}
{\bf D}(J)&=&\frac 1{2\pi }4\int_0^1\left( b_{-}dx\right) \;\left(
b_{-}^2x^2\right) \sqrt{2E\left( 1-k^2x^2\right) \left( 1-x^2\right) }
\nonumber\\
&=&\frac 2\pi \sqrt{2E}b_{-}^3\sigma \left[ k\right],  \label{functiond}
\end{eqnarray}
where
\begin{equation}
\sigma \left[ k\right] =\int_0^1dx\;x^2\sqrt{\left( 1-k^2x^2\right) \left(
1-x^2\right) }.  
\label{sigma}
\end{equation}

Following a suggestion in WW 22.72, this can be reduced to complete elliptic
integrals of the first and second kinds (we will need the third kind for the
$S$ function), to get the result quoted in the main text.

The function  ${\bf S}$ is given by
\[
{\bf S}=\frac{-1}{4\pi ^2}\int_0^Tdt\;\left[ \frac d{dt}b^2(t)\right]
{\rm Pf}\int_0^\infty \frac{du}u\;b^2(t-u) .
\]
Let us consider an orbit beginning at $b\left( 0\right) =0$, and divide the
time interval in four quarters:
I) $0\leq t\leq T/4$,
II) $T/4\leq t\leq T/2$,
III) $T/2\leq t\leq 3T/4$,
IV) $3T/4\leq t\leq T$.
We have the following relationships:
I) in the first quarter, $b_I=b\left( t\right) ,$ $p_I=p\left( t\right) $,
II) in the second quarter, $b_{II}(t)=b_I\left( \frac T2-t\right) $,
$p_{II}=-p_I\left( \frac T2-t\right) $,
III) in the third quarter, $b_{III}(t)=-b_I\left( t-\frac T2\right) $,
$p_{III}=-p_I\left( t-\frac T2\right) $,
IV) in the fourth quarter, $b_{IV}(t)=-b_I\left( T-t\right) $,
$p_{II}=p_I\left( T-t\right) $.
This suggests parametrizing time in terms of a unique variable $\tau$,
$0\leq \tau \leq T/4$, as follows:
I) In the first quarter, $t=\tau$,
II) In the second quarter, $t=T/2-\tau$,
III) In the third quarter, $t=T/2+\tau$,
IV) In the fourth quarter, $t=T-\tau$.

We can then write
\begin{eqnarray*}
{\bf S} &=&\frac{-1}{2\pi ^2}\int_0^{T/4}d\tau \;b(\tau )p\left( \tau
\right) {\rm Pf}\int_0^\infty \frac{du}u\; \\
&&\left[ b^2(\tau -u)-b^2\left( \frac T2-\tau -u\right) +b^2\left( 
\frac T2+\tau -u\right) -b^2\left( T-\tau -u\right) \right].
\end{eqnarray*}
Since $b^2$ is an even function of $t$ with period $T/2$, we have
\[
{\bf S}=\frac{-1}{\pi ^2}\int_0^{T/4}d\tau \;b(\tau )p\left( \tau \right)
{\rm Pf}\int_0^\infty \frac{du}u\left[ b^2(\tau -u)-b^2\left( \tau +u\right)
\right] , 
\]
and since the second integrand is obviously even
\begin{equation}
{\bf S}=\frac 1{\pi ^2}\int_0^{T/4}d\tau \;b(\tau )p\left( \tau \right)
{\rm Pf}\int_{-\infty }^\infty \frac{du}ub^2\left( \tau +u\right).  
\label{reduceds}
\end{equation}

To proceed, we must appeal to the addition theorem for elliptic functions
(AS 16.17.1). Next we use the differential equation for Jacobi elliptic
functions (AS 16.16.1) and integrate by parts to get
\begin{equation}
{\bf S}=\frac{32EE_s}{\pi ^2}k^2{\rm Pf}\int_0^\infty 
\frac{du}{u^2}{\rm sn}^2\left( u\right) \rho \left[ k^2{\rm sn}^2\left(
u\right) \right], \label{readys}
\end{equation}
where
\begin{equation}
\rho \left[ n\right] =\int_0^1dx\;\frac{x^2\sqrt{\left( 1-k^2x^2\right)
\left( 1-x^2\right) }}{\left[ 1-nx^2\right] },  \label{readyr}
\end{equation}
which can be expressed in terms of complete elliptic integrals
\begin{equation}
\rho =\left( -k^2\right) \left\{ \left[ c^{\prime }+\frac a{k^2}\right]
K\left[ k\right] -\frac a{k^2}E\left[ k\right] -c\Pi \left[ n,k\right]
\right\},  \label{newr}
\end{equation}
where the last term is the complete elliptic integral of the third kind (AS
17.7.2), with $\sin \alpha =k$, 
\begin{eqnarray}
a&=&\frac 1n\left[ \frac 1n-\frac{\left( 1+k^2\right) }{3k^2}\right],
\nonumber\\
c^{\prime }&=&\frac 1n\left[ \frac 2{3k^2}+\frac 1n\left( \frac 1n-
\frac{\left( 1+k^2\right) }{k^2}\right) \right], 
\nonumber\\
c&=&\frac 1n\left[ \frac 1{k^2}+\frac 1n\left( \frac 1n-\frac{\left(
1+k^2\right) }{k^2}\right) \right].
\nonumber
\end{eqnarray}
Since in our application we allways have $n\leq k^2$, we may use formulae AS
(17.7.6) and (17. 4.28) to get the result in the text (recall that
$E/E_s=4k^2/\left( 1+k^2\right) ^2$)

\subsection{Relaxation time}

The aim of this section is to estimate the time on which a solution to
the transport equation with arbitrary initial conditions relaxes to a steady
solution as discussed in the main body of the paper, in section \ref{section4}.
The way this kind of problem is usually handled \cite{Kurchan 1998} is to write
the Fokker-Planck equation (\ref{FokkerPlanck}) in a way ressembling a
(Euclidean) Schr\"odinger equation

\begin{equation}
\frac{\partial f}{\partial t}=Lf  \label{transport}.
\end{equation}
Then if a complete basis of eigenfunctions of the $L$ operator can be found
\begin{equation}
Lf_n\left( p,q\right) =E_nf_n\left( p,q\right),  \label{right}
\end{equation}
a generic solution to Eq. (\ref{transport}) reads
\begin{equation}
f(p,q,t)=\sum c_nf_n\left( p,q\right) e^{E_nt}.  \label{solution}
\end{equation}
Therefore, provided no eigenvalue has a positive real part, the relaxation
time is the inverse of the real part of the largest nonzero eigenvalue. The
$L$ operator may have purely imaginary eigenvalues, in which case it does not
relax towards any steady solution.

This problem differs from the ordinary quantum mechanical one in several
aspects, the most important being that the $L$ operator does not have to be
either Hermitian or anti-Hermitian. That is why the eigenvalues will be
generally complex, rather than just real or imaginary. Also, it is important
to notice that the ``right'' eigenvalue problem Eq. (\ref{right}) is
different from the ``left'' eigenvalue problem:
$ g_n\overleftarrow{L}=E_n^{\prime }g_n$. 
For example, for any $L$ of the form $L=\partial _iK^i,$ where the $K$'s are
themselves operators, $g_0\equiv 1$ is a solution to this (left) equation 
(with zero eigenvalue), while it may not be a solution to Eq. (\ref{right}) at
all.

\subsubsection{Our problem}

In our case, the $L$ operator can be read from Eq. (\ref{FokkerPlanck}).
Since we are taking $\nu $ as a small parameter, it is natural to write
$ L=L^0+L^1 $, where
\begin{eqnarray}
L^0f&=&\left\{ H,f\right\},  \label{ele0}\\
L^1f&=&-\frac \partial {\partial p}\left[ Ff\right] 
+\frac{\pi \nu ^2\Lambda ^2}4b^2\frac{\partial ^2f}{\partial p^2}.  
\label{ele1}
\end{eqnarray}

The spectral decomposition of $L^0$ is very simple. In action-angle
variables 
\begin{equation}
L^0f=-\Omega \left( J\right) \frac{\partial f}{\partial \theta }.
\label{ele02}
\end{equation}
Imposing periodicity in $\theta $ we find the following eigenvalues: $0$ and
\begin{equation}
E_{n,\chi }^0=-in\Omega \left( \chi \right),  \label{zerothavl}
\end{equation}
with $n$ integer (note that $L^0$ is anti-Hermitian). The eigenvalue $0$ is
infinitely degenerate: any function of $J$ alone is an eigenvector with zero
eigenvalue. The $E_{n,\chi }^0$ have eigenfunctions
\begin{equation}
f_{n,\chi }^0\left( J,\theta \right) =\frac{e^{in\theta }}
{\sqrt{2\pi }}\delta \left( J-\chi \right),  \label{efe-n-chi}
\end{equation}
and, barring accidental degeneracy (the ratio of frequencies for two
different actions being rational) are non degenerate. These eigenfunctions are
normalized with the Hilbert product
$\left( g\left| f\right. \right) =\int_0^{J_s}\int_0^{2\pi} 
dJd\theta \;g^{*}f$ as
$\left( 0n\xi \left| 0n\chi \right. \right) =\delta 
\left( \xi -\chi \right) $, where here and in the rest of this section we use
Dirac's notation.

Having solved the eigenvalue problem for $L^0$, it is only natural to see
that of $L$ as an exercise in time independent perturbation theory. There
are three differences with the ordinary textbook problem:
1) $L^1$ is neither Hermitian nor anti-Hermitian;
2) one of the eigenvalues of $L^0$ is degenerate;
3) the eigenfunctions of $L^0$ are not normalizable.
In spite of this, the basic routine from quantum mechanics textbooks still
works.

\subsubsection{Perturbations to nonzero eigenvalues}

Let us seek the first order correction to $E_{n,\chi }^0$. We write the
exact eigenvalue as
$ E_{n,\chi }=E_{n,\chi }^0+E_{n,\chi }^1+... $
corresponding to the exact eigenfunction
$f_{n,\chi }=f_{n,\chi }^0+f_{n,\chi }^1+... $, 
and obtain
\begin{equation}
L^1f_{n,\chi }^0+L^0f_{n,\chi }^1=E_{n,\chi }^1f_{n,\chi }^0+
E_{n,\chi}^0f_{n,\chi }^1 .  \label{linpert}
\end{equation}

For $m\neq n$ we multiply both sides of the equation by $f_{m,\xi }^{0*}$, use
that $L^0$ is anti-Hermitian and integrate over $J$ and $\theta $,
to get
\begin{equation}
\left( 0m\xi \left| 1n\chi \right. \right) =\frac{\left( 0m\xi \left|
L^1\right| 0n\chi \right) }{E_{n,\chi }^0-E_{m,\xi }^0}. 
\end{equation}

In the $m=n\neq 0$ case, the same operation yields
\begin{equation}
E_{n,\chi }^1\left( 0n\xi \left| 0n\chi \right. \right) =\left( 0n\xi \left|
L^1\right| 0n\chi \right) -\left[ E_{n,\chi }^0-E_{n,\xi }^0\right] \left(
0n\xi \left| 1n\chi \right. \right) ,
\end{equation}
and we may write
\begin{equation}
L^1f_{n,\chi }^0=\frac{e^{in\theta }}{\sqrt{2\pi }}\left[ R+iI\right], 
\end{equation}
where
\begin{equation}
R=L^1\delta \left( J-\xi \right) -n^2\frac{\pi \nu ^2\Lambda ^2}4b^2\left(
\left. \frac{\partial \theta }{\partial p}\right| _b\right) ^2\delta \left(
J-\xi \right).  \label{real}
\end{equation}
Whatever the imaginary part $I$ is, it is not relevant to the relaxation
time; in a similar way, the average of the first term in Eq. (\ref{real})
yields no term proportional to $\left( 0n\xi \left| 0n\chi \right. \right)$
Therefore, we conclude that
\begin{equation}
{\rm Re}\left[ E_{n,\chi }^1\right] =-n^2\frac{\pi \nu ^2\Lambda ^2}
{8\pi }\left. \int_0^{2\pi} d\theta \;b^2\left( \left. \frac{\partial \theta }
{\partial p} \right| _b\right) ^2\right| _{J=\chi }.  \label{reltime1}
\end{equation}
We see on dimensional grounds alone that the relaxation time (the inverse of
this equation) will be of order $E_s^2$ (recall that $E_s=3/(2\Lambda)$, much
shorter than the average tunneling time, which is proportional to the inverse
of (\ref{tunb}).

The expression (\ref{reltime1}) may be slightly simplified by using the
identity $(\partial \theta/\partial p)|_b=-
(\partial b/\partial J)|_\theta$,
which follows from the transformation from one set of variables to the other
being canonical. We may write
\begin{equation}
{\rm Re}\left[ E_{n,\chi }^1\right] =-n^2\frac{\pi \nu ^2\Lambda ^2}{32\pi }
\left. \int_0^{2\pi} d\theta \;\left( \left. \frac{\partial b^2}{\partial
J}\right| _\theta \right) ^2\right| _{J=\chi }. \label{reltime2}
\end{equation}
We may Fourier transform $b^2$ as a function of $\theta $, derive term by
term, and use Parseval's identity, to conclude that in any case
\begin{equation}
\left| {\rm Re}\left[ E_{n,\chi }^1\right] \right| \geq n^2\frac{\pi \nu
^2\Lambda ^2}{\left( 8\pi \right) ^2}\left. \left[ \frac d{dJ}\int_0^{2\pi}
d\theta \;b^2\right] ^2\right| _{J=\chi }.  \label{reltime3}
\end{equation}
The integral in this expression can be performed, 
recall that $b=b_- x(t)$ where $x(t)$ is given in
(\ref{elliptic}). We recall also that $\Omega=\theta t$ with $\Omega$ given in
(\ref{frequency}), and then use as integration variable $2\theta K[k]/\pi$,
where $K[k]$ is the elliptic integral defined in (\ref{new3}), to get finally
\begin{equation}
\left| {\rm Re}\left[ E_{n,\chi }^1\right] \right| \geq n^2\frac{\pi \nu
^2\Lambda ^2}{\left( 4\pi \right) }\left[ \frac d{dJ}\left( \frac{b_{-}^2}
{k^2}\left[ 1-\frac{E\left[ k\right] }{K\left[ k\right] }\right] \right)
\right] ^2 .
\end{equation}

Rather than a general formula, let us investigate the limiting cases.
For $J\rightarrow 0$, we have: $J\sim E $, $k^2\sim E/(4E_s)$,
$b_{-}^2\sim 2E$, 
$E[ k] \sim (\pi/ 2)( 1-k^2/4)$, and
$K[ k] \sim (\pi/ 2)( 1+k^2/4)$. In this limit
we thus get
\begin{equation}
\left| {\rm Re}\left[ E_{n,\chi }^1\right] \right| \geq n^2\frac{
\nu^2\Lambda ^2}{4}. 
\end{equation}

For $J\rightarrow J_s$ (near the separatrix) we can use the following
approximations: $b_{-}^2\sim 4E_s\left[ 1-\sqrt{1-E/E_s}\right] $,
$k^2\sim \left[ 1-2\sqrt{1-E/E_s}\right] $,
$K[ k] \sim (1/2)\ln [16/(1-k^2)]\sim (1/4)\ln
[64/(1-E/E_s)]$, 
$E[ k] \sim 1+(1/4)\sqrt{1-E/E_s}\{  \ln
[64/(1-E/E_s)]-1\}$, and  
$dE/dJ=\Omega \sim  \pi /(2K[ k])$. Thus 
the correction to the eigenvalue diverges. In both cases, we get that
the relaxation time is much smaller than the tunneling time.

\subsubsection{Perturbation of the zero eigenvalue}

We now confront the harder problem of finding the first order correction to
the zero eigenvalue. The idea, as in quantum mechanics, is that
the first order eigenvalues shall be the eigenvalues of the restriction of
$L^1$ to the proper subspace of the zero eigenvalue, namely, the infinite
dimensional space of all $\theta $ independent functions.
If $f_{0,\chi }$ corresponds to an eigenfunction with null eigenvalue, the
first order secular equation becomes
\begin{equation}
L^1f_{0,\chi }^0+L^0f_{0,\chi }^1=E_{0,\chi }^1f_{0,\chi }^0.
\label{linpert2}
\end{equation}
We elliminate the second term in the left hand side of this equation by
projecting back on $\theta $ independent functions, by averaging over
$\theta $. Fortunately the average over $\theta $ of $L^1$ acting on a $\theta $
independent function is precisely what we did in section \ref{section3}, so
using (\ref{b6}) and (\ref{b10}) we can write down the eigenvalue problem
\begin{equation}
\frac{\pi \nu ^2\Lambda ^2}4\frac d{dJ}\left[ \frac {\bf D}\Omega \frac d{dJ}
-{\bf S}\right] f=\lambda f , \label{secular}
\end{equation}
where we call $\lambda $ the eigenvalue, to avoid confussion with the energy.
The left hand side of this equation is a sum of two terms, the first one being
Hermitian, and the second undefined. However, if we introduce a new function
$\Psi$ by
$f=\Psi \exp \left[ \frac 12\int^EdE^{\prime }\;\beta \left( E^{\prime}
\right) \right]$ where $\beta={\bf S}/{\bf D}$ we can write
\begin{equation}
\frac{\pi \nu ^2\Lambda ^2}4\left\{ \frac d{dJ}\frac {\bf D}\Omega \frac d{dJ}
-\frac 12\frac{d{\bf S}}{dJ}-\frac{\Omega {\bf S}^2}{4{\bf D}}\right\} \Psi
=\lambda \Psi.\label{secular2}
\end{equation}
Recall that we have seen in section \ref{section3} that ${\bf S}$ is an
increasing function of $E$ (or $J$). Therefore, multiplying by $\Psi ^{*}$ and
integrating, we see that $\lambda $ must be real and negative. This is an
important result.

Let us introduce a new non negative parameter $\alpha$,
\begin{equation}
\lambda =-\frac{\pi \nu ^2\Lambda ^2}8\alpha,  \label{lambda2alfa}
\end{equation}
and write equation (\ref{secular2}) using $E$ as independent variable instead
of $ J $ ($dE/dJ=\Omega$), and then introduce a new function $\psi$ by
$\Psi=\psi/\sqrt{{\bf D}}$. Finally (\ref{secular2}) becomes 
\begin{equation}
-\frac 12\psi ^{\prime \prime }+V_\alpha \left( E\right) \psi =0
\label{effschrod}
\end{equation}
where
\begin{equation}
V_\alpha \left( E\right) =\frac 1{4{\bf D}}\left\{ \frac{d{\bf S}}{dE}
+\frac{{\bf S}^2}{2{\bf D}} +{\bf D}^{\prime \prime }-\frac{{\bf D}^{\prime
2}}{2{\bf D}}-\frac \alpha \Omega \right\}, \label{effpot} 
\end{equation}
which looks like a Schr\"odinger equation with a weird potential.
We have therefore transformed the problem of finding the eigenvalues of
equation (\ref{secular2}) into the question of for which values of $\alpha $
a particle of zero energy has a bound state in the potential $V_\alpha
\left( E\right)$.

To get an idea of what is going on, let us make the approximation ${\bf D}\sim
cE^2 $, ${\bf S}\sim \beta {\bf D}$, where $c$ and $\beta $ are constant, then
\begin{equation}
V_\alpha \left( E\right) =\frac \beta {4E^2}\left[ 2E+\frac \beta 2E^2-
\frac \alpha {c\beta \Omega }\right]  \label{effpot2}
\end{equation}

When $\alpha =0$, we should get back some results of section
\ref{section4}. Indeed, in this case the solutions for large $E$ go like $\exp
(\pm\beta E/2)$, which, after the equation relating $f$ with $\Psi$, means that
the solutions either are exponentially growing or bounded. The first ones
correspond to steady solutions with non zero flux (those in section
\ref{section4}), while the second ones are the stationary solutions with no flux.
Note that the change from $\Psi$ to $\psi$, that we made previously, enforces
the pathological $E^{-1}$ low energy behavior we found in section \ref{section4}.

For $\alpha \neq 0$, the effective potential $V_\alpha $ has two classical
turning points, i.e. points where $V_\alpha(E)=0$. For small $E$ we find 
$E_1\sim  \alpha /(2c\beta ) $ (we use that $\Omega(E_1)\sim 1$), and for
large $E$ we find $E_2$ given by
$\Omega ^{-1}\left( E_2\right) \sim c\beta ^2E_s^2/(2\alpha) $,
which under the asymptotic form
$\Omega^{-1} ( E) \sim \ln [64/(1-E/E_s)]/(\sqrt{2}\pi)$,
is
$E_2\sim E_s\{ 1-64\exp [ -\pi c\beta ^2 E_s^2/(\sqrt{2}\alpha)]\}$. 
The first classically allowed region sits precisely where the theory is
unreliable, and we ought to disregard it as an artifact. Therefore the low
$\alpha $ eigenstates must be related to the presence of the second allowed
region, near the separatrix. This is consistent with the fact that the
zeroth order eigenvalues are $-in\Omega $ (see Eq. (\ref{zerothavl})), and
so they tend to accumulate around $0$ as we approach the separatrix.

In the second classically allowed region (large $E$) we may approximate
\begin{equation}
V_\alpha \left( E\right) \sim\frac \alpha {4cE_s^2}\left[ \frac 1{\Omega \left(
E_2\right) }-\frac 1{\Omega \left( E\right) }\right].  \label{effpot3}
\end{equation}
As an estimate, we may look for values of $\alpha $ such
as $V_\alpha $ satisfies a Bohr-Sommerfeld condition
\begin{equation}
\int_{E_2}^{E_s}dE\;\sqrt{-2V_\alpha \left( E\right) }\sim n\pi,  \label{BS}
\end{equation}
(this only makes sense if we treat the separatrix as a turning point). To
perform the integral, we introduce a new variable
$x=\ln [(1-E_2/E_s)/(1-E/E_s)]$.
The integral turns out to be
$n\pi \sim \sqrt{\alpha}( 1-E_2/E_s)
\int_0^\infty dx\;\sqrt{x}e^{-x}/\sqrt{2\sqrt{2}\pi c}$, 
and so the eigenvalues are the roots of
\begin{equation}
\alpha _n\exp \left( -\frac{\sqrt{2}\pi c\beta ^2E_s^2}{\alpha _n}\right)=
\frac{n^2\pi ^2c}{128}. \label{quant}
\end{equation}
The relevant value of $c$ being $0.96$ near the separatrix, see the end of
section \ref{section3}, thus $\beta \sim 1.23$. Taking the log of Eq.
(\ref{quant}), we find the lowest eigenvalue
\begin{equation}
\alpha _1=\frac{\sqrt{2}\pi c\beta ^2E_s^2}{\ln \left( 128
\sqrt{2}\beta^2 E_s^2/\pi \right) }\left[ 1+O\left( \frac{\ln \ln E_s} {\ln
E_s}\right) \right].  \label{lowestavl}
\end{equation}

This is the result we were looking for. Going back to
the beginning, we translate this into eigenvalues of the Fokker-Planck
operator, see Eqs. (\ref{right}) and (\ref{secular}), 
\begin{equation}
\lambda \sim -\frac{9\pi \nu ^2}{32}\frac{\sqrt{2}\pi c\beta ^2}{\ln \left( 
128 \sqrt{2}\beta ^2E_s^2/\pi\right) },  \label{decayrate}
\end{equation}
where we have used (\ref{lambda2alfa}) and that $ E_s=3/(2\Lambda)$. Thus
we conclude that the relaxation time grows logarithmically with $E_s$, while
the tunneling time grows exponentially. In fact, the tunneling time is
proportional to the inverse of (\ref{tunb}), and so it goes like 
$\sim\exp(1.23\, E_s)$. Therefore it is totally justified to analyze tunneling
under the assumption that all transient solution have died out, and we only have
the steady solutions discussed in section \ref{section4}.

\subsection{A single cosmic cycle}

The purpose of this section is to discuss whether it is possible to generalize
the discussion of the paper to models with a single cosmic cycle. The basic
problem is that an universe emerging from the singularity with a finite
expansion rate is bound to lead to infinite particle production 
\cite{Calzetta and Castagnino 1984}. Therefore, in
order to make sense, it is unavoidable to modify the behavior of the model close
to the singularity, and there is no unique way to do this. Of course, a
possibility is to assume that the singularity behaves as a perfectly reflecting
boundary, which is equivalent to what we have done so far. Another
possibility, to be discussed here, is that the evolution is modified for
very smal universes, so that $p$ vanishes as $b\rightarrow 0$. For example,
if the initial stages of expansion (and the final stages of collapse) are
replaced by an inflationary (deflationary) period, then $p\sim b^2,$ $\dot p
\sim b^3$, etc. We shall assume such an evolution in what follows. In these
models, the singularity is literally pushed to the edge of time.

\subsubsection{The {\bf D} and {\bf S} functions}

The ${\bf D}$ function is given by Eq. (\ref{b2}), where now we
average over a half period only. However, the periodicity of the integrand
is precisely $T/2$, so the average over a half period is the same as the
full average. Therefore, ${\bf D}\sim E^2/2$ at low energy, and $0.96\,E^2$
close to the separatrix as we had in the many cycles model.

For the function ${\bf S}$, let us begin from Eq. (\ref{b4}),
modified to represent average over a half period
\begin{equation}
S\left( J\right) =\frac 1\pi \int_0^{T/2}dt\;pF(b,p,t),  \label{alfa}
\end{equation}
then use Eq. (\ref{51}) for $F$ and integrate by parts twice to get
\begin{equation}
S\left( J\right) =\frac{6\nu }\pi \left[ p\frac d{dt}\frac Ib\right]
_0^{T/2}-\frac{6\nu }\pi \left[ \dot p\frac Ib\right] _0^{T/2}+
\frac{6\nu } \pi \int_0^{T/2}dt\;\left( \frac{\ddot p}b-
\frac{p\ddot b}{b^2}\right) I(b,p,t).  \label{delta}
\end{equation}

The discussion above on the approach to the singularity means that the
integrated terms vanish. In the remaining term we use the equations of motion
$\ddot b=\dot p=-V^{\prime }\left( b\right)$ and get
\begin{equation}
S\left( J\right) =\frac{\nu \Lambda }\pi \int_0^{T/2}dt\;\frac{db^2}{dt}
I(b,p,t).  \label{epsilon}
\end{equation}
Next use Eqs. (\ref{52}), (\ref{53d}), (\ref{48}), (\ref{50}) and the
redefinition (\ref{b10}) to write
\begin{equation}
{\bf S}=-\frac 1{2\pi ^2}\int_0^{T/2}dt\;\frac{db^2}{dt}
{\rm Pf}\int_0^t\frac{du}u \;b^2\left( t-u\right), \label{kappa}
\end{equation}
where we have truncated the $u$ integral to restrict it to the range  where
the equations of motion hold.

Instead of looking for a general expression, we shall only consider the
low energy limit and the behavior close to the separatrix.

\subsubsection{Low energy limit}

For low energy,
$b=(\sqrt{2E}/\Omega)\sin \Omega t $. Substituting this into Eq.
(\ref{kappa}), changing the order of integration and performing some simple
integrations we obtain,
\begin{equation}
{\bf S=}\frac{E^2}{2\pi ^2\Omega ^4}{\rm Pf}\int_0^{T/2}
\frac{du}u\;\left[ 1-\cos 2\Omega u+\pi \sin 2\Omega u\right]=
6.89 \;\frac{E^2}{2\pi^2\Omega^4},  \label{low}
\end{equation}
where the last integration has been performed numerically.
Thus,  ${\bf S}$ retains the main
features as in the previous case, the most important being the sign and energy
dependence.

\subsubsection{Close to the separatrix}

Close to the separatrix, we must make allowance for the fact that the orbit
spends an increasing amount of time near the turning point $b_{-}$. It is
thus convenient to isolate the central portion of the orbit.
Let us rewrite Eq. (\ref{kappa}) as
\begin{equation}
{\bf S}=-\frac 1{2\pi ^2}\left[ \int_0^{T/4}dt\;\frac{db^2}{dt}
\int_0^t\frac{du}u\;b^2\left( t-u\right) +\int_{T/4}^{T/2}dt\;
\frac{db^2}{dt}\int_0^t\frac{ du}u\;b^2\left( t-u\right) \right].  
\label{xi}
\end{equation}
Divide the $u$ integral by quarter orbits,
write $t=T/2-t^{\prime }$ in some of these integrals, and use the
periodicity and parity of $b^2$ and $db^2/dt$.
We can then rewrite ${\bf S}$ as
\begin{equation}
{\bf S}=A+B  \label{zeta}
\end{equation}
where
\begin{eqnarray}
A&=&-\frac 1{2\pi ^2}\left[ \int_0^{T/4}dt\;\frac{db^2}{dt}\int_0^{T/4+t}
\frac{du}u\;b^2\left( t-u\right) 
-\int_0^{T/4}dt\;\frac{db^2}{dt}\int_0^{T/4-t}
\frac{du}u\;b^2\left( t+u\right) \right],  \nonumber\\
B&=&\frac 1{2\pi ^2}\left[ \int_0^{T/4}dt\;\frac{db^2}{dt}\int_t^{T/4+t}
\frac{du}u\;b^2\left( t-u\right) +\int_0^{T/4}dt\;\frac{db^2}{dt}
\int_{T/4-t}^{T/2-t}\frac{du}u\;b^2\left( t+u\right) \right].  \nonumber
\end{eqnarray}
Observe that the factor $db^2/dt$ effectively cuts off the $t$ integrals at
times much shorter than $T/4$. So we can take the limit $T\rightarrow \infty$,
whereby $A$ converges to the expression for ${\bf S}$ of the previous case,
i.e. Eq. (\ref{reduceds}). Here, our problem is to estimate $B$.

Let us write $B=C+D$, where
\begin{eqnarray}
C&=&\frac 1{2\pi ^2}\int_0^{T/4}dt\;\frac{db^2}{dt}\int_0^{T/4}
\frac{dv}{t+v} \;b^2\left( v\right), \nonumber \\
D&=&\frac 1{2\pi ^2}\int_0^{T/4}dt\;\frac{db^2}{dt}\int_0^{T/4}
\frac{dv}{T/2-v-t}\;b^2\left( v\right). \nonumber
\end{eqnarray}
To evaluate $C$, we integrate by parts
and take the limit $T\rightarrow \infty $,
\begin{equation}
C=\frac{b_{-}^4}{2\pi ^2}\ln \left( \frac T4\right) -\frac 1{2\pi ^2}
\int_0^\infty dt\;\frac{db^2}{dt}\int_0^\infty dv\;\ln \left( t+v\right) \;
\frac{db^2}{dv}+O\left( \frac 1T\right).  \label{eta}
\end{equation}
Let us use the same argument in $D$, take the limit and add $C$ to get $B$.
The final result is
\begin{equation}
{\bf S}=\frac 1{2\pi ^2}\left\{ \int_0^\infty dt\;\frac{db^2}{dt}
\int_0^\infty du\;\left[ \frac 1u\left\{ b^2\left( t+u\right) -\;b^2\left(
t-u\right) \right\} -\ln \left( t+u\right) \;\frac{db^2}{du}\right]
+b_{-}^4\ln \left( \frac T2\right) \right\}.  \label{theta}
\end{equation}
Using that at the separatrix  $b=\sqrt{4E_s}\tanh (t/\sqrt{2})$, the double
integral in the above expression gives $13.89\, E_s^2/(2\pi^2)$, and we finally
have, 
\begin{equation}
{\bf S}=0.70\, E_s^2+\frac{8E_s^2}{\pi ^2}\ln \left( \frac
T2\right).
\end{equation}
For $T$ we have the result (cfr. Eq. (\ref{frequency}))
$T=4\sqrt{2}K[k]/(1+\sqrt{1-E/E_s}) $,
and when $k\rightarrow 1$,
$K[ k] \sim (1/4)\ln [64/(1-E/E_s)] $,
and ${\bf S}$ can then be written as,
\begin{equation}
{\bf S}=0.42 \,E_s^2+\frac{8E_s^2}{\pi ^2}\ln \left( \ln \frac{64}{1-E/E_s}
\right).  \label{finalS}
\end{equation}

\subsubsection{The flux}

We shall now show that, in spite of the divergence in ${\bf S}$, $f$ itself
remains finite as we approach the separatrix. Basically, the arguments in
section \ref{section4} still hold, so the equation to solve is
\begin{equation}
\frac{df}{dE}-\left[ \beta +\alpha \ln \left( \ln \frac{64}{1-E/E_s}
\right) \right] f=0,
\end{equation}
where $\beta =0.44$ ($0.42/0.96$) and $\alpha =0.84$ ($8/0.96\pi^2$).
Let us now call $64\,e^{-x}=1-E/E_s$, then $dE=64\,E_se^{-x}dx$ and the equation
becomes,
\begin{equation}
\frac{df}{dx}-64\,E_s\left[ \beta +\alpha \ln x\right] e^{-x}f=0 , \label{newEq}
\end{equation}
which is well behaved as $x\rightarrow \infty $.

In order to estimate the flux, we now need the integral of $f$ in a
neighborhood of the separatrix, namely
$K^{-1}\sim \int dE\;f $.
With the same change of variables as above, we get
\begin{equation}
K^{-1}\sim 64E_s\int^\infty dx\;\exp \left[ 64E_s\int^xdx^{\prime }\;\left(
\beta +\alpha \ln x^{\prime }\right) e^{-x^{\prime }}-x\right].
\label{newflux}
\end{equation}
The integral peaks when
$64\,E_s( \beta +\alpha \ln x) e^{-x}=1 $, which defines
$x_0=\ln (64\,E_s)+\ln ( \beta +\alpha \ln x_0) $, and thus
\begin{equation}
K^{-1}\sim \frac 1{\beta +\alpha \ln x_0}\exp \left[
64\,E_s\int^{x_0}dx^{\prime }\;\left( \beta +\alpha \ln x^{\prime }\right)
e^{-x^{\prime }}\right]. 
\end{equation}
In order to get back the old result when $\alpha =0$, we must assume a lower
limit for the integral at $x\sim \ln 64\sim 4.16$, which corresponds to 
$E\sim 0$. This limit is high enough that the integral is dominated by the lower
limit ($e^{-x}\ln x$ peaks below $e$), so we finally obtain,
\begin{equation}
K\sim (\beta +\alpha \ln x_0)\exp\left[- ( \beta +1.62\alpha) E_s\right]
\sim ({\rm prefactor})\exp \left(-\frac {2.71}{\Lambda}\right).
\label{finalflux}
\end{equation}

This result should be compared to our previous result (\ref{tunb}), or
(\ref{b24}). In spite of everything, we are still above the quantum
tunneling probability (\ref{b26}). Thus, considering a cosmological model
which undergoes a single cosmic cycle does not qualitatively change our
conclusions.

\end{document}